\documentclass[12pt]{article}
\usepackage{amsmath}
\usepackage[utf8]{inputenc}
\usepackage[left=3.5cm, right=3.5cm, top=3cm]{geometry}
\usepackage{natbib}
\usepackage{graphicx}
\usepackage{amssymb}
\usepackage{booktabs}
\usepackage{caption}
\usepackage{subcaption}
\usepackage{subfloat}
\usepackage{bm} 
\usepackage[dvipsnames]{xcolor}
\usepackage[onehalfspacing]{setspace}
\usepackage{float}
\usepackage{url}

\providecommand{\keywords}[1]
{
	\small	
	\textbf{\textit{Keywords: }} #1
}

\begin{document}

\begin{titlepage}

	\Huge
	\centering
	\textbf{Separating the Wheat from the Chaff: \\Bayesian Regularization in Dynamic Social Networks} \\
	\vspace{2cm}
	
	\large
	Diana Karimova$^a$, Roger Th.A.J. Leenders$^{b,c}$, Joris Mulder$^{a,b}$

\small 

\begin{flushleft}
		$^a$ Department of Methodology and Statistics, Tilburg School of Social and Behavioral Sciences, Tilburg University\\
$^b$ Jheronimus Academy of Data Science \\
$^c$ Department of Organization Studies, Tilburg School of Social and Behavioral Sciences, Tilburg University	
\end{flushleft} 

\vspace{2cm}

\begin{flushleft}
\textbf{Corresponding author: \\} 
Diana Karimova \\
\textbf{Email:} d.karimova@uvt.nl\\
\textbf{Work Address:} Warandelaan 2, Tilburg, Netherlands, 5037 AB

\end{flushleft}

\end{titlepage}

\begin{center}
	\large
	\textbf{Separating the Wheat from the Chaff: \\Bayesian Regularization in Dynamic Social Networks} \\
	\vspace{1cm}
\end{center}

\begin{abstract}
In recent years there has been an increasing interest in the use of relational event models for dynamic social network analysis. The basis of these models is the concept of an ``event", defined as a triplet of time, sender, and receiver of some social interaction. The key question that relational event models aim to answer is what drives social interactions among actors. Researchers often consider a very large number of predictors in their studies (including exogenous variables, endogenous network effects, and  various interaction effects). The problem is however that employing an excessive number of effects may lead to model overfitting and inflated Type-I error rates. Consequently, the fitted model can easily become overly complex and the implied social interaction behavior becomes difficult to interpret. A potential solution to this problem is to apply Bayesian regularization using shrinkage priors. In this paper, we propose Bayesian regularization methods for relational event models using four different priors: a flat prior model with no shrinkage effect, a ridge estimator with a normal prior, a Bayesian lasso with a Laplace prior, and a horseshoe estimator with a numerically constructed prior that has an asymptote at zero. We develop and use these models for both an actor-oriented relational event model and a dyad-oriented relational event model. We show how to apply Bayesian regularization methods for these models and provide insights about which method works best and guidelines how to apply them in practice. Our results show that shrinkage priors can reduce Type-I errors while keeping reasonably high predictive performance and yielding parsimonious models to explain social network behavior.
    
\end{abstract}	

\keywords{Bayesian regularization, shrinkage priors, Bayesian lasso, \\ horseshoe prior,  relational event data}
	
\section{Introduction}	

Relational event history data is becoming increasingly available, partly due to increased use of technology-supported interaction, such as email, phone calls, and online social networks (Twitter, Facebook, et cetera). Relational event data encode who does what with respect to whom at what point in time. Typically, relational event data contain information of the exact timing (or order) of interactions, who was the sender, who was the receiver, and, possibly, what was the mode of communication (e.g., face-to-face or digital), the sentiment (e.g., positive or negative), the content, et cetera. Additionally, information about attributes of the involved actors is also often available, for example gender, group memberships, or the hierarchical position of an actor. Finally, information about possible external influences to the event history, such as deadlines, the start of new projects, or time-of day/day-of-week effects in in social networks in organizations, may be available. Thus, due to its high-resolution, relational event data can potentially greatly support our understanding of complex interaction processes, about temporal effects of interventions in networks, or about how past interactions may affect what will happen in the (nearby) future.

Relational event models (initially proposed by \cite{butts20084}, and later extended by  \cite{dubois2013hierarchical}, \cite{quintane2014modeling}, \cite{leenders2016once}, \cite{pilny2016illustration},  \cite{vu2017relational}, \cite{stadtfeld2017dynamic}, \cite{mulder2019modeling},  \cite{lerner2020reliability}, among others) have become widely used over the past decade for analyzing relational event data. In the relational event model, the outcome variable is the rate of interaction between potential senders and receivers in the network at a given time point. This rate of interaction is commonly assumed to be a log linear function of a set of predictor variables (at that given point in time). Following \cite{leenders2016once}, predictor variables can be categorized as being endogenous (summarizing the past activity of the actors in the network) or exogenous (capturing actor attributes or external influences). Essentially, the relational event model aims to predict the next event: given the past interactions among the actors and the characteristics of the actors, at a given time \textit{t}, what will be the next interaction (and when will it occur)? In relational event models, this is achieved by finding a loglinear combination of endogenous and exogenous predictors that best (or, sufficiently) model the rates of interaction of every event that is possible at \textit{t}.

There are at least three ways in which researchers aim to build relational event models (and these approaches are also frequently applied in the fitting of other complex network models for which substantive theory is scant, such as (S)(T)ERGM's or SIENA-models). First, when researchers have a theory of what drives the social interaction dynamics in a network, predictors are chosen to reflect the theory at hand. Unfortunately, there is almost no successful theory that explains who interacts with whom and at what rate (and who don't), so there is rarely a decent theoretical foundation for researchers to base the selection of predictors on. A second approach is to engage in a data mining-like approach by adding and removing predictors until the best achievable model has been reached. Typically, researchers start with a set of predictors (inspired by theory or previous findings), remove those that are not statistically significant and add new ones that might improve accuracy further, until the researcher is satisfied with the model fit. This is a potentially time-consuming approach that makes the interpretation of \textit{t}-statistics highly questionable and disallows the model to be used for inferential purposes. A third approach is to throw a great many variables into the model and then interpret the signs and statistical significance of all of the predictors. This latter approach is statistically more acceptable, but the researcher runs the risk of including statistics that are highly mutually correlated and the resulting model may very well overfit the data. Although we are much in favor of the first approach, it is undermined by the dearth of solid theory to begin with. In an empirical project, the researcher therefore needs an approach to separate the wheat from the chaff in order to distinguish which predictors really matter and which do not. In statistics, this can be be addressed using variable selection algorithms.

Variable selection algorithms have not yet been thoroughly developed for relational event models, but researchers have developed some ways to help the variable selection process. For example, \cite{butts20084} proposed a model for explaining radio communication messages between emergency transponders during the 9/11 World Trade Center disaster. To make decisions about which variables to include, \cite{butts20084} utilized the BIC, a model selection criterion that balances model fit and model complexity (via the number of predictor variables). In most cases, however, it is not computationally feasible to compute the BIC for all possible models since the more predictors are considered, the more models need to be fitted in order to compute all of the BIC's and find the model with the lowest BIC. Hence, in practice researchers only compare a few competing models. This choice of which models to compare is inherently arbitrary (to an extent) considering the aforementioned lack of solid theory to build on, and may be driven by computational burden (the longer it takes for a model to run, the fewer models can feasibly be compared).

Another possible solution for variable selection problems exists in a form of penalized or regularized regression. In penalized regression a penalty term is added to the sum of squared residuals. This results in shrinkage of small negligible effects (moving them to zero) while leaving large effects largely unaffected. Therefore, small effects are removed from the model (as the coefficients end up close to zero and may lose statistical significance) and the effects that truly matter stay intact. The result is a more parsimonious model that highlights which effects really matter, without the ``clutter" of many effects that do not contribute much statistically. Both frequentist and Bayesian regularization approaches have been shown to effectively guard against overfitting and to result in good predictive performance \citep{tibshirani1996regression, park2008bayesian, kyung2010penalized, van_erp2019shrinkage}. In this paper, we therefore contend that a useful and statistically sound alternative to the third approach above would be to specify a fairly large model (i.e. include all potential predictors of interest, regardless of multicollinearity) and then use regularization to separate the wheat from the chaff among all of the predictors at once.

In particular, we introduce a Bayesian regularization approach for relational event models. One of the benefits of Bayesian approach is that Bayesian regularization performs competitively and sometimes better than its classical counter parts (in terms of \textit{predictive mean squared error}) and results in more accurate standard errors (via posterior standard deviations) \citep{park2008bayesian}. Furthermore, a Bayesian approach is natural for regularization since the prior naturally serves as a penalty function. Bayesian approaches are not restricted to non-concave penalty functions that are computationally difficult to optimize in a maximum likelihood framework. Another advantage of the Bayesian approach is the possibility to fit all parameters of the model in one step, instead of using ad-hoc two-step methods where the penalty parameter needs to be separately determined, e.g., using cross-validation. In this paper, we will consider four different priors that each regularize relational event models differently. First, we will consider a flat prior to  serve as a reference as it results in no shrinkage. This makes it comparable to ordinary maximum likelihood. Second, we consider Bayesian ridge regression using a Gaussian prior. Third, we develop Bayesian lasso regression using LaPlace priors \citep{park2008bayesian}, which has thicker tails than the Gaussian prior and a higher peak at zero. Finally, we consider the (non-concave) horseshoe prior that is constructed via an F distribution. Originally proposed by \cite{carvalho2010horseshoe}, the horseshoe prior has a spike at zero and thicker tails than a Cauchy distribution. We will discuss the implementation of these Bayesian regularization algorithms and illustrate their use for both a dyadic and an actor-oriented relational event model.

In Section 2 of this paper we explain two partial likelihood approaches to relational event modelling: one for a dyadic relational events and one for  actor-oriented events. In Section 3 we discuss the four priors for Bayesian regularization, and in Section 4 we illustrate the methodology with two empirical examples. We conclude and provide final discussions in Section 5.

\section{Relational event modelling using partial likelihoods}

The relational event model was popularized by \cite{butts20084}, who modeled a relational event history as a Poisson process where the event rate for each specific dyad depends on a set of endogenous and exogenous effects through a loglinear function.

Instead of working with the full likelihood for a joint model for all event times, senders, and receivers, in this paper we adopt the idea of partial likelihoods \citep{cox1972regression,perry2013point} that only considers specific parts of a conditional likelihood. By working with partial likelihoods, we simplify the specification of the model by focusing on the outcome variables that are of most interest for a given application. Below, we first present a partial likelihood that is actor-oriented, was proposed by \cite{perry2013point} and further developed by \cite{stadtfeld2017dynamic}. This approach may be preferred when a researcher is mainly interested in modeling the choice of the receiver of an event conditional on the sender \citep{vu2017relational, stadtfeld2017interactions, hoffman2020model, hedstrom2009analytical}. Second, we provide a partial likelihood for a dyadic relational event model which has not yet been considered in the literature, to our knowledge. This latter approach may be preferred when one is interested in modeling the full dyad (sender and receiver jointly) (\cite{leenders2016once}, \cite{brandes2009networks}, and numerous direct applications of REM, such as \cite{malang2019networks}, \cite{liang2014organizational}, \cite{lerner2018let}.

\subsection{Partial likelihood for an actor-oriented \\model}

Using the notation of events $e_m = (t_m, s_m, r_m), \, m \in \{1, \dots, M\} $ as a triplets of time $t$, sender $s$, and receiver $r$, we can write the likelihood of the sequence of events as a product of conditional likelihoods: 

\begin{equation}
	\begin{split}
		L(e_1, ..., e_M) =& L(e_1)L(e_2,..., e_M) = L(e_1)L(e_2|e_1) \cdot\ldots\cdot L(e_M|e_1,\ldots,e_{M-1})  = \\
		=& L(t_1, s_1)L(r_1|t_1, s_1)L(t_2, s_2|t_1, s_1, r_1)L(r_2|t_2, s_2, e_1)\cdot ...\cdot \\
		\cdot & L(t_M, s_M|e_{M-1}, ..., e_1)L(r_M|t_M, s_M, e_{M-1}, ... , e_1) \\
	\end{split}
	\label{L_all}
\end{equation}

Without loss of generality, we can focus on the choice of the receiver, for a given point in time and a given sender. In most research projects, understanding who will be the receiver of an event is more informative than modeling who will be the sender. We thus consider the following partial likelihood of the receivers of the events conditional on the senders and event times, which follows directly from equation \eqref{L_all}: 
\begin{equation}
	\begin{split}
		PL(\textbf{r} | \textbf{s}, \textbf{t}) =& L(r_1|t_1, s_1)\cdot L(r_2|t_2, s_2) \cdot \dots \cdot L(r_M|t_M, s_M)\\
	\end{split}
	\label{PL_receiver}
\end{equation}

The partial likelihood in \eqref{PL_receiver} can be seen as a statistical choice model, where the sender ``chooses" the most suitable receiver of an interaction from the set of possible receivers. In this paper we consider a Bayesian probit model using a Gaussian latent variable approach by extending the work of \cite{imai2005bayesian} to relational data. For each event $e_i$ in a sequence $\{e_1,..., e_M \},$ where M is a total number of events, we define a categorical outcome variable $Y_i$ that represents a receiver of the event $e_i$. This receiver of event $i$ can be any actor in the \textit{risk set} $\mathcal{R}_{actor}$ -- the set of actors who are possible receivers of a given event. For simplicity, we assume that all actors, except for the sender, are at risk. In our latent variable approach the sender assigns a latent ``suitability scale" to all potential receivers in the risk set. The latent suitability of actor $r$ for a given sender $s_i$ is denoted by $Z_{ir}$. The receiver $r$ with the largest $Z_{ir}$ will be the predicted receiver of the event:

\begin{equation}
Y_i(Z_i) = r, \mbox{ if } \max(Z_i) = Z_{ir},
\label{eq:latent}
\end{equation}
where $Z_i = (Z_{i1}, \dots, Z_{iN})$ is a multivariate latent variable. In the framework of the multivariate probit model, we can write 
\begin{equation}
Z_i = X_{i} \bm{\beta} + \epsilon_i,
\label{regression_actor}
\end{equation}

where $X_i$ is a $N \times P$ matrix of observed predictor variables at time $i$, $\bm{\beta} = (\beta_1, ..., \beta_P)^T$ is a vector of network parameters, and  $\epsilon_i$ is a Gaussian error term, centered at zero, having an identity covariance matrix (to ensure identifiability of the model). The matrix $X_i, i = 1,..., M$ of predictor variables can include endogenous as well as exogenous predictors, defined for each actor in the risk set. Network effects that are caclulated on a pair of sender and receiver are often not defined for loop events, when a sender and a receiver are the same actor. In such cases the corresponding row of matrix $X_i$ is empty.

\subsection{A partial likelihood for a dyadic model}

In certain applications the interest is in jointly modeling the combination of sender and receiver \citep{leenders2016once,brandes2009networks,malang2019networks, liang2014organizational, lerner2018let}. Thus, unlike the choice model of the receiver given the sender, we build a statistical model for all possible dyads that can be observed at a given point in time. Starting from the same full likelihood, but redefining the conditional likelihood in such a way that we condition on the time points of events, we get the following representation of the likelihood:

\begin{equation}
	\begin{split}
		L(e_1, ..., e_M) =& L(e_1)L(e_2,..., e_M) = L(e_1)L(e_2|e_1) \cdot\ldots\cdot L(e_M|e_1,\ldots,e_{M-1})  = ... = \\
		=& L(t_1)L(s_1, r_1|t_1)L(t_2|t_1, s_1, r_1)L(s_2, r_2|t_2, e_1)\cdot ...\cdot \\
		\cdot & L(t_M|e_{M-1}, ..., e_1)L(s_M, r_M|t_M, e_{M-1}, ... , e_1) \\
	\end{split}
	\label{L_all2}
\end{equation}

A partial likelihood for a dyad REM can be written as follows: 

\begin{equation}
	\begin{split}
		PL(\textbf{s},\textbf{r} | \textbf{t}) =& L(s_1, r_1|t_1)\cdot L(s_2, r_2|t_2) \cdot \dots \cdot L(s_M, r_M|t_M)\\
	\end{split}
	\label{PL_dyad}
\end{equation}
This can be viewed as a dyadic partial likelihood for the REM where the sender and receiver for event $e_i$ are jointly modeled.

In contrast with the previous model, in the dyadic model the outcome variable is the rate of a dyad. In particular, $Y_i$ is defined as an index of the dyad $(s, r)$ from the risk set $\mathcal{R}_{dyad}$ of all the $N(N-1)$ possible ordered dyads. Following the idea of a multivariate probit model, we assume that all dyads that are at risk lie on a ``latent activity scale"; the dyad with the largest latent activity score becomes the dyad that is predicted to occur next:
\[
Y_i(W_i) = \textit{l}(s_i, r_i), \mbox{if } \max(W_i) = \textit{l}(s_i, r_i),
\]
where $\textit{l}(s_i, r_i)\in \{1, \dots, N(N-1)\} $ is the index of dyad $(s_i, r_i)$ in ordered risk set $\mathcal{R}_{dyad}$. Therefore, latent vectors $W_{i} $ have length $N(N-1)$, under the condition that an actor cannot send an event to oneself. Note that a long length of the latent variable can lead to a high computational cost for larger networks as there will be many unknown latent variables to be estimated.  

We write the regression for the dyad REM as follows: 
\begin{equation}
	W_i = X_{i} \bm{\beta} + \epsilon_i, \, i = 1, ..., M
	\label{regression_dyad}
\end{equation}
where matrices $X_i$ are of dimension $N(N-1)\times P$ (containing the dyadic predictor variables) and $\bm\beta$ is the corresponding vector that quantifies the relative importance of the predictors. As in the actor-oriented model, the set of potentially important dyadic predictor variables can be huge. To find the true nonzero effects, a Bayesian regularization algorithm will be proposed, as discussed in the next section.

\section{Bayesian regularization via shrinkage priors}\label{bayesreg}

The complete Bayesian model combines the statistical model for the data with a prior distribution for the network parameters $\bm\beta$\footnote{In this section we describe the priors in the case of actor-oriented model. The priors for the dyadic model are mathematically equivalent to the ones described in sections \ref{flat}, \ref{ridge}, \ref{lasso}, \ref{horseshoe}.}:
\begin{eqnarray*}
	\text{model}&:&\left\{
	\begin{array}{ccl}
		Y_i(Z_i) &=& j, \mbox{ if } \max(Z_i) = Z_{ij}, \\
		Z_i &=& X_i \bm\beta + \epsilon_i, \text{ with } \epsilon_i \sim N(0,I)
	\end{array}
	\right.\\
	\text{prior} &:& p(\bm\beta)
\end{eqnarray*}

In many applications of Bayesian statistical methods, the prior is used to add external information to the analysis, e.g., reflecting expert knowledge or previous empirical findings. In Bayesian regularization, on the other hand, the prior acts as a penalty function that shrinks negligible effects towards zero while leaving large effects mostly unaffected. This is why the priors in these models are often referred to as ``shrinkage priors." The result of this behavior is that the resulting model is more parsimonious than the unregularized model, showing which variables matter a lot and which might as well be neglected. To allow the same type of shrinkage behaviour for negative and positive effects, the prior should have a symmetric form with a peak at zero (to shrink small effects) while having some probability mass allocated in the tails (to leave large effects largely unaffected).

\begin{figure}
	\centering
	\includegraphics[width=0.5\linewidth]{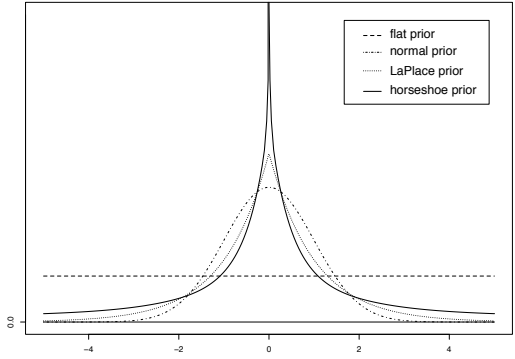}
	\caption{Flat prior (dashed line), normal prior (dash-dotted line), LaPlace prior (dotted line), and horseshoe prior (solid line).}
	\label{fig:priors}
\end{figure}

Different types of priors can be used for this purpose. For a recent overview, see \cite{van_erp2019shrinkage}. In the current paper, we consider three of the most popular shrinkage priors from the Bayesian regularization literature: a Gaussian prior (for Bayesian ridge regression), a LaPlace prior (for Bayesian lasso regression), and a horseshoe prior. Figure \ref{fig:priors} displays these priors. We add a flat (horizontal) improper prior that does not perform any shrinkage (as it assumes that all values of the parameters are equally likely a priori). The results of the flat prior are comparable with maximum likelihood estimation, making it a perfect reference.

The variance of a shrinkage prior has a direct effect on the amount of shrinkage in the model: a large (small) prior variance induces little (considerable) shrinkage. This prior variance is controlled via the \textit{shrinkage parameter} and is denoted as $\lambda^2$. Ideally, when there are many large effects, the shrinkage parameter should be large (so the large effects are left intact), and when there are hardly any large effects, the shrinkage parameter should be small (so the small effects are nudged towards zero). The optimal value of the shrinkage parameter for a given data set can be found using two-step approaches such as cross-validation or empirical Bayes methods \citep[e.g.,][]{park2008bayesian}. A more natural choice is to estimate the shrinkage parameter jointly with the Bayesian model in one step, yielding a full Bayesian model. This only requires one to specify a separate prior density for the shrinkage parameter $\lambda^2$.

To fit the Bayesian regularized relational event models, we use Markov Chain Monte Carlo (MCMC) methods to sample the parameters from the joint posterior. The approach is to sample the model parameters sequentially from their conditional posterior distributions. Gibbs sampling is a specific MCMC algorithm where the conditional posterior distributions of the parameters belong to known distributional families from which it is easy to sample. This is generally the case when the priors have the same distributional form as the likelihood (this is known as ``conjugacy"). Because of the Gaussian errors of the latent variables in the partial likelihoods from Section 2, Gaussian priors for $\bm\beta$ result in conditional posteriors that also have Gaussian distributions. It is possible to write the Bayesian lasso prior and the horseshoe prior conditionally as scaled mixture of Gaussian priors, to ensure easy and efficient posterior sampling using Gausian posteriors. For the shrinkage parameters, we consider $F$ priors as they are relatively vague while allowing easy posterior sampling using Gamma and Inverse distributions \citep{mulder2018matrix}. This is equivalent to choosing a half-Cauchy prior for $\lambda$, which is a common choice \citep{carvalho2009handling}.  In the next section, we discuss each Bayesian shrinkage model for relational event analysis in detail.

\subsection{Flat prior (no shrinkage)}\label{flat}

We first consider a benchmark model with no shrinkage effect. This model utilizes a flat improper prior that assumes that all values for the regression coefficients vector $\bm{\beta} = (\beta_1, \dots, \beta_P)$ are equally likely a priori. Mathematically, this can be written as 
\begin{equation}
	p^{FLAT}(\bm\beta) \propto 1,
\label{flat_prior}	
\end{equation}

The prior density is constant over the complete real line (see the dashed line in figure \ref{fig:priors}). The prior does not shrink the regression coefficients: the estimates of the $\bm{\beta}$ will be entirely driven by the data. The Bayesian flat prior model, therefore, behaves very similar to classical MLE estimation. Given an observed event history we can estimate the posterior distribution of $\bm{\beta} = (\beta_1, \ldots, \beta_P)$. Because the latent variable has a multivariate normal distribution, the model can be estimated using a Gibbs sampler that can be written as:

\begin{enumerate}
	\item Set initial values for $\bm\beta^{(0)}$, and $Z^{(0)}$.
	\item Given $Z^{(s-1)}$, draw $\bm{\beta}^{(s)}$ from its conditional posterior distribution that follows the multivariate normal distribution:

	$$\bm\beta^{(s)} | Z^{(s-1)} \sim N(\mu_{\beta}, \Sigma_{\beta}), \mbox{where} $$ 
	
	$$\mu_{\beta} = \left( \sum\limits_{i=1}^{M} X_i^T X_i\right)^{-1}\sum\limits_{i=1}^{M} X_i^T Z_i^{(s-1)} \mbox{,  } \Sigma_{\beta} = \left( \sum\limits_{i=1}^{N} X_i^T X_i\right)^{-1} $$
	
	\item Given $\bm\beta^{(s)}$, draw $Z_{ir}^{(s)}$ from its conditional posterior distribution that follows the truncated normal distribution:
	
	$$ 
	Z_{i}^{(s)} | \bm\beta^{(s)} \sim t\mathcal{N}(X_i\bm{\beta}^{(s)}, I_N).
	$$
	To reduce the degrees of freedom when choosing the latent variables, the first component of each $Z_i$ is fixed to zero. The value of the element of $Z_i$ that corresponds to the observed actor is generated from a truncated density on the interval $(\max_{r\not = r_i} Z_{ir},\infty)$. Alternatively, the values of the elements that correspond to actors that are not observed are generated from the truncated density on the opposite interval $(-\infty,Z_{ir_i})$. This way we can guarantee that latent variables resemble the given relational event data.
	\item Repeat steps 2 and 3 for $s=1,\ldots,S$.
\end{enumerate}

The initial set of draws is discarded as they are part of the burn-in period and would depend on the arbitrarily chosen initial values. The remainder is used to construct Bayesian credible intervals, or point estimates such as posterior mode, for each $\beta_1, ..., \beta_P$. These intervals indicate whether the data suggest that the effect is negligible or not. 

\subsection{Bayesian ridge prior}\label{ridge}

Ridge regression was originally developed to improve estimates of the classic least squares model, especially in the case when there is a high correlation among predictors. The model utilizes a modified variance matrix $X'X+\lambda^2 I$ that adds a quadratic penalty. In Bayesian ridge regression, a normal prior is used for the regression coefficients:
\begin{equation}
	p^{RIDGE}(\bm\beta | \lambda^2) = \prod_{p=1}^P p(\beta_p | \lambda^2) = \prod_{p=1}^P
	\mathcal{N}(\beta_p | 0,\lambda^2),
\end{equation}
where $\mathcal{N}(\beta_p | 0,\lambda^2)$ denotes a normal prior for $\beta_p$ with mean 0 and variance $\lambda^2$.

We plot this prior in Figure \ref{fig:priors} using a dash-dotted line. The prior is centered around zero with relatively thin tails. Shrinkage is performed over the entire domain of parameters due to the structure of normal prior density: large values will be shrunk to the same degree as are small values.

To complete the Bayesian model, we need a prior for the shrinkage parameter $\lambda^2$. A common prior for this purpose is the gamma distribution (\cite{park2008bayesian}). However, the hyperparameters of the gamma prior may considerably affect its results \citep{kyung2010penalized}. For this reason, we use a half-Cauchy prior for $\lambda$ instead, which is quite vague due to its thick tails. A half-Cauchy prior for $\lambda$ is equivalent to a matrix-$F$ prior for $\lambda^2$ \citep[e.g.,][]{mulder2018matrix}, which has density: 
\begin{equation}
	F(\lambda^2; \alpha_1,\alpha_2,b) = \frac{\Gamma(\frac{\alpha_1+\alpha_2}{2})}{\Gamma(\frac{\alpha_1}{2})\Gamma(\frac{\alpha_2}{2})}b^{-\alpha_2/2}\left(\frac{\lambda^2}{b} + 1 \right)^{-\frac{\alpha_1+\alpha_2}{2}} (\lambda^2)^{\alpha_2/2-1}
	\label{F_density_alpha}
\end{equation}
The hyperparameters will be set to 1, which is the default minimally informative choice:
\[
\lambda^2 \sim F(1,1,1)
\]

Using a parameter expansion, the matrix-$F$ distribution can be written as a scale mixture of inverse gamma distributions via
\[
F(\lambda^2| \alpha_1,\alpha_2,b) = \int IG(\lambda^2|\tfrac{\alpha_2}{2},\delta)G(\delta|\tfrac{\alpha_1}{2},b)d\psi^2.
\]
This makes the prior conditionally conjugate, which makes the MCMC algorithm quite efficient. The Gibbs Sampler algorithm is as follows: 

\begin{enumerate}
	\item Set initial values for $\bm\beta^{(0)}$, and $Z^{(0)}$, as well as for parameters $\lambda^{2(0)}$, $\delta^{(0)}$.
	\item Draw $\bm{\beta}^{(s)}$ from its conditional posterior distribution given $Z^{(s-1)}$, 

	$$\bm\beta^{(s)} | Z^{(s-1)} \sim \mathcal{N}(\mu^{\textit{ridge}}, \Sigma^{\textit{ridge}}), \mbox{ where} $$

	$$ \mu^{\textit{ridge}} = \left( \sum\limits_{i=1}^{M} X_i^T X_i + \frac{1}{\lambda^2} I_P \right)^{-1} \sum_{i=1}^{M} X_i^T Z_i^{(s-1)}$$
	
	$$\Sigma^{\textit{ridge}} = \left( \sum\limits_{i=1}^{M} X_i^T X_i + \frac{1}{\lambda^{2(s-1)}} I_P \right)^{-1} $$

	\item Draw $Z_{ir}^{(s)}$ from its conditional posterior given $\bm\beta^{(s)},$ which is a truncated normal distribution.
	
	$$ 
	Z^{(s)} | \bm\beta^{(s)} \sim t\mathcal{N}(X_i\bm{\beta}^{(s)}, I_N)
	$$ 
	
	To ensure that the sampled values of the latent variables correspond to the given data and satisfy equation (\ref{eq:latent}), we sample latent variables in the following way: first element of each vector $Z_i$ is set to zero to reduce degrees of freedom; the element that corresponds to the observed actor is sampled from a truncated normal density on the interval $(\max_{r\not = r_i} Z_{ir},\infty)$; and the elements which correspond to non-observed actors are sampled from truncated density on the complement interval $(-\infty,Z_{ir_i})$.
	
	\item Draw the shrinkage parameter $\lambda^2$ and the expanded parameter $\delta$:
    \begin{eqnarray*}
	 \lambda^{2(s)}|\bm{\beta}^{(s)}, \delta^{(s-1)} &\sim& \mbox{IG}(\alpha_1+\frac{P}{2}, \delta^{(s-1)} +\frac{1}{2}\sum\limits_{p=1}^{P}(\beta^{(s)}_p)^2)
	 \\
	  \delta^{(s)}|\lambda^{2(s)} &\sim& \mbox{G}(\alpha_1+\alpha_2, \frac{1}{\lambda^{2(s)}}+\frac{1}{b})
	 \end{eqnarray*}
	\item Repeat steps 2 to 4 for $s=1,\ldots,S$.
\end{enumerate}

After discarding the initial set of draws as a burn-in period, the remainder is used to construct Bayesian credibility intervals.

\subsection{Bayesian lasso prior}\label{lasso}

The classical lasso (``least absolute shrinkage and selection operator'') regression model uses a $L_1$ norm as a penalty term, which is the sum of the absolute values of the regression coefficients. The Bayesian equivalent of the lasso penalty is obtained by using a Laplace prior for regression coefficients (\cite{park2008bayesian}).

\begin{equation}
	p^{LASSO}(\bm\beta | \lambda^2) = \prod_{p=1}^P Laplace(\beta_p | \lambda^2), \\
\end{equation}
for $p$ = $1,\ldots, P$. To facilitate Bayesian computation, the Laplace prior can be written as a normal distribution where the scale has an exponential distribution; this results in a conditionally conjugate Bayesian model:
 \[
Laplace(\beta_p | \lambda^2) = \int \mathcal{N}(\beta_p | 0,\tau_p^2\lambda^2)Exp(\tau_p^2|1) d\tau_p.
\]

We plot the Laplace prior as a dotted line in Figure \ref{fig:priors}; it is more peaked around zero compared to the normal (ridge) prior. In opposite to the ridge model estimates, this results in stronger shrinkage for small estimated effects and, due to the Laplace having thicker tails than the normal prior, less shrinkage for larger estimated effects.

Compared to the normal (ridge) prior, the lasso prior includes the parameter $\tau_p^2$; this serves as a shrinkage parameter on a local level for effect $\beta_p$ and varies across the $\beta_p$. The $\lambda^2$ parameter, on the other hand, controls global shrinkage and affects all $\beta_p$ to the same degree. The idea of a separate global and a local shrinkage parameter was introduced by \cite{carvalho2009handling} and allows a researcher to control the shrinkage behavior of the method precisely. To complete the model, we set a vague matrix-$F$ prior for the global shrinkage parameter:
\[
\lambda^2 \sim F(1,1,1).\\
\]

The steps in the Gibbs sampler are the following: 
\begin{enumerate}
	\item Set initial values for $\bm{\beta}^{(0)}, Z^{(0)}, \lambda^{2(0)}, \tau^{2(0)}_1, \dots,\tau^{2(0)}_P, \delta^{(0)} $
	\item  Draw $\bm{\beta}^{(s)}$  from its conditional posterior distribution given $Z^{(s-1)}, \tau_1^{2(s-1)}, ..., \tau_P^{2(s-1)}$, $\lambda^{2(s-1)}$
	
    $$	 \bm{\beta}^{(s)}| Z^{(s-1)}, \tau_1^{2(s-1)}, ..., \tau_P^{2(s-1)}, \lambda^{2(s-1)} \sim \mathcal{N}(\mu^{\textit{lasso}}, \Sigma^{\textit{lasso}}), \mbox{ where }$$
    
    $$ \mu^{\textit{lasso}} = \left(\sum\limits_{i=1}^{M}X_i^TX_i + D_{\tau}^{-1} \right)^{-1}\sum\limits_{i=1}^{M}X_i^TZ_i^{(s-1)} $$ 
	
	$$ \Sigma^{\textit{lasso}} = \left(\sum\limits_{i=1}^{M}X_i^TX_i + D_{\tau}^{-1} \right)^{-1},  $$

	 $$D_{\tau} = diag\{\lambda^{2(s-1)} \tau_1^{2(s-1)}, \dots, \lambda^{2(s-1)} \tau_P^{2(s-1)}\}$$
	
	\item Update latent variables by sampling $Z_{ir}^{(s)}$ from its conditional posterior given $\bm\beta^{(s)},$ which is a truncated normal distribution, such that 
	
	$$ 
	Z^{(s)} | \bm\beta^{(s)} \sim t\mathcal{N}(X_i\bm{\beta}^{(s)}, I_N).
	$$ 
	and for an element of $Z_i$ that corresponds to the observed actor the truncated interval is $(\max_{r\not = r_i} Z_{ir},\infty)$ while elements that conform the actors that are not observed are truncated in the interval $(-\infty,Z_{ir_i})$. These conditions will guarantee that sampled latent variables fit the observed categorical data and according to equation (\ref{eq:latent}). In addition, the first element of each $Z_i$ is fixed to zero, such that there are less degrees of freedom in the generating process.

	\item Draw the value of parameter $\bm{\tau}^{2(s)} = (\tau^{2(s)}_1, \dots, \tau^{2(s)}_P)$ from its conditional posterior distribution, which in this case is an inverse-Gaussian distribution:
	
	$$\frac{1}{\tau_p^{2(s)}}| \beta_p^{(s-1)}, \lambda^{2(s-1)} \sim \mbox{Inv-Gauss}(\mu'= \sqrt{\frac{\lambda^{2(s-1)}}{\beta_p^{2(s-1)}}}, \lambda' = 1 ), p = 1, \dots, P$$
	
	\item Update the values of $\lambda^{2(s)}$ (similarly to the step 4 for the ridge model): 
	
	$$ \lambda^{2(s)}| \tau^{2(s-1)}_1, ..., \tau^{2(s-1)}_P, \delta^{(s-1)} \sim IG(\alpha_1+\frac{P}{2}, \delta + \frac{1}{2} \sum\limits_{p=1}^{P} \frac{\beta^2_p}{\tau^2_p})$$
	
    $$p(\delta^{(s)}| \lambda^{2(s-1)}) \propto G(\alpha_1 + \alpha_2, \frac{1}{\lambda^{2(s-1)}} + \frac{1}{b})$$
    
    \item Repeat steps 2 to 5 for $s=1,\ldots,S$.
	
\end{enumerate}

After discarding the burnin period, the posterior draws can be used to obtain posterior estimates and construct Bayesian credibility intervals.

\subsection{Bayesian horseshoe prior}\label{horseshoe}
The horseshoe model was first introduced in \cite{carvalho2010horseshoe} and has the key shape characteristics of an asymptote at zero and heavier tails. To construct this prior, the original model proposes to use a half-Cauchy distribution (i.e., a Student $t$ distribution with 1 degree of freedom). This results in heavier shrinkage of small effects and less shrinkage of large effects compared to the Bayesian lasso. The prior can be written as follows:

\begin{equation}
	p^{HORSESHOE}(\bm\beta | {\lambda}^2) = \prod_{p=1}^P Horseshoe(\beta_p | \lambda^2),\\
\end{equation}
where $\lambda^2$ is a global shrinkage parameter. Again, to facilitate Bayesian computation the horseshoe prior can be written as a scaled mixture of normals where $\lambda^2$ follows a matrix-$F$ distribution (which is equivalent with a half-Cauchy prior for $\lambda$), i.e.,

\[
Horseshoe(\beta_p | \lambda^2) = \int
	\mathcal{N}(\beta_p | 0,\lambda^2\tau_p^2)F(\tau_p^2|1,1,1)d\tau_p^2.
\]

A graphical representation of the horseshoe prior is given in Figure \ref{fig:priors} by the solid line. As can be seen the prior has a sharp peak at zero and heavy tails. The name ``horseshoe" comes from the observation that the shrinkage coefficient $\kappa_p$, defined as $\kappa_p = 1/(1+\tau_p^2)$, has a horseshoe-shaped $Beta(1/2, 1/2)$ distribution under the matrix-$F$ prior for $\tau_p$. This coefficient reflects the amount of weight that the posterior places around zero: $\kappa_p \approx 0$ corresponds to no shrinkage whereas $\kappa_p \approx 1$ corresponds to total shrinkage to zero. 

 Similar to the Bayesian lasso prior, $\tau^2_p$ serves as a local shrinkage parameter for $\beta_p$, while $\lambda^2$ serves as a global shrinkage parameter. In contrast to the Bayesian lasso, the local shrinkage parameters now follows an $F$ distribution instead of an exponential distribution. As the $F$ distribution has thicker tails than the exponential in case of Bayesian lasso, the $F$ prior will result in less shrinkage of large effects than the lasso model. The parameter $\lambda^2$ optimizes the overall level of sparsity, while the local shrinkage parameters $\tau_p^2$ prevent large effects from being shrunk towards zero. Again, we finalize the Bayesian horseshoe model by setting a vague $F$ prior on the global shrinkage parameter:
 \[
 \lambda^2 \sim F(1,1,1).
 \]
 
 The model can be fitted efficiently using a Gibbs sampler algorithm with the following steps: 
\begin{enumerate}
	\item Set initial values for $\bm{\beta}^{(0)}, Z^{(0)}, \bm{\lambda}^{2(0)} =(\tau^{2(0)}_1, \dots, \tau^{2(0)}_P), \lambda^{2(0)},$ and mixing parameters of the $F$ densities:  $ \bm{\psi^{2(0)}} =  (\psi^{2(0)}_1, \dots, \psi^{2(0)}_P), \gamma^{2(0)}.$

	\item  Draw $\bm{\beta}^{(s)}$  from its conditional posterior distribution given $Z^{(s-1)}, \bm{\tau}^{2(s-1)}, \lambda^{2(s-1)}$
	
	$$	 \bm{\beta}^{(s)}| Z^{(s-1)}, \bm{\tau}^{2(s-1)}, \lambda^{2(s)} \sim \mathcal{N}(\mu^{\textit{horseshoe}}, \Sigma^{\textit{horseshoe}}), \mbox{ where }$$
	
	$$ \mu^{\textit{horseshoe}} = \left(\sum\limits_{i=1}^{M}X_i^TX_i + D_{\tau}^{-1} \right)^{-1}\sum\limits_{i=1}^{M}X_i^TZ_i^{(s-1)} $$ 
	
	$$ \Sigma^{\textit{horseshoe}} = \left(\sum\limits_{i=1}^{M}X_i^TX_i + D_{\tau}^{-1} \right)^{-1},  $$
	
	$$D_{\tau} = diag\{\lambda^{2(s-1)} \tau_1^{2(s-1)}, \dots, \lambda^{2(s-1)} \tau_P^{2(s-1)}\}$$
	
	\item Draw the values of latent variables by sampling $Z_{ir}^{(s)}$ from conditional posterior given $\bm\beta^{(s)},$ which is a truncated normal distribution
	
	$$ 
	Z^{(s)} | \bm\beta^{(s)} \sim t\mathcal{N}(X_i\bm{\beta}^{(s)}, I_N).
	$$ 
	
	under the conditions that for an observed element of $Z_i$ the truncated interval is $(\max_{r\not = r_i} Z_{ir},\infty)$ and for elements that are not observed the truncated interval is $(-\infty,Z_{ir_i})$, while the first element of $Z_i$ is always set to zero. Such conditions form latent variables that correspond to the observed categorical data by the the definition of equation (\ref{eq:latent}).
	
	\item Draw the value of parameter $\bm{\tau}^{2(s)}$ from its conditional posterior - inverse gamma distribution
	
	 $$\tau_p^{2(s)}| \psi_p^{2(s-1)}, \lambda^{2(s-1)}, \beta_p^{(s)} \sim \mbox{IG}\left( \alpha_3 + \frac{1}{2},\psi_p^{2(s-1)} +\frac{\beta_p^{2(s)}}{2\lambda^{2(s-1)}}\right) , \, p=1, \dots, P$$

	\item Draw the value of parameter $\lambda^{2(s)}$ from its conditional posterior - inverse gamma distribution
	
	$$\lambda^{2(s)}| \gamma^{2(s-1)}, \bm{\tau}^{2(s)}, \bm{\beta}^{(s)} \sim \mbox{IG}\left( \alpha_1 + \frac{P}{2}, \gamma^{2(s-1)} + \frac{1}{2}\sum\limits_{p=1}^{P}\frac{\beta^{2(s)}_p}{\tau^{2(s)}_p}\right) $$
	
	\item Update the mixing parameters $\bm{\psi^{2(s)}}$ and $\gamma^{2(s)}:$ 
	
	$$\gamma^{2(s)}|\lambda^{2(s)} \sim \mbox{G}\left( \alpha_1 + \alpha_2, \frac{1}{b_1}+\frac{1}{\lambda^{2(s)}} \right) $$

    $$\psi_p^{s(s)}| \tau^{2(s)}_p \sim \mbox{G} \left(  \alpha_3+\alpha_4, \frac{1}{b_2} + \frac{1}{\tau_p^{2(s)}} \right) , p = 1, \dots, P  $$

	\item Repeat steps 2 to 6 for $s=1,\ldots,S$.

\end{enumerate}

\subsection{Comparison of priors in a simple hypothetical case}

In order to illustrate the shrinkage effect of the four models we estimate the shrinkage models on relational event sequences of fixed length across increasing effect sizes. For this illustration we consider the actor-oriented model (but the shrinkage behavior is similar for the dyadic model). We create event sequences with different effect sizes by properly specifying the design matrix X. For example, consider a sequence of 30 relational events on a network of six actors, assuming a scalar network parameter $\beta$ and a single predictor variable that is zero for all actors except for actor 2 (for whom it is equal to $0.2$) for all events in the sequence $(i \in \{1, \dots, 30\})$:

\[ X_i =  \begin{pmatrix}
	0 \\ 0.2 \\ 0 \\ 0 \\ 0 \\ 0  \\ 
\end{pmatrix}  
\]

Without the loss of generality, we assume that Actor 1 is the sender of all events. The number of actors who receive an event from actor 1 decreases across the sequences as shown in Table \ref{seq}. In Sequence 1 all events are sent proportionally to all actors 2,3,$\cdots$, 6. In the last Sequence 29 all events but one (to avoid identification issues) are sent to actor 2. 

\begin{table}	
\[\begin{matrix}
	\mbox{Event} & \mbox{Sending} & \mbox{Receiver} & \mbox{Receiver} & \mbox{Receiver} &  & \mbox{Receiver}    \\
	\mbox{index} & \mbox{actor} & \mbox{Sequence 1} & \mbox{Sequence 2} & \mbox{Sequence 3} & \cdots & \mbox{Sequence 29}    \\
	1& 1& \textbf{2} & \textbf{2} & \textbf{2} & & \textbf{2}\\
	2& 1& 3 & \textbf{2} & \textbf{2} & & \textbf{2}\\
	3& 1& 4 & 4 & \textbf{2} & & \textbf{2}\\
	4& 1& 5 & 5 & 5 & & \textbf{2}\\
	5& 1& 6 & 6 & 6 & & \textbf{2}\\
	6& 1& \textbf{2} & \textbf{2} & \textbf{2} & & \textbf{2}\\
	7& 1& 3 & 3 & 3 & & \textbf{2}\\
	8& 1& 4 & 4 & 4 & & \textbf{2}\\
	9& 1& 5 & 5 & 5 & & \textbf{2}\\
	10& 1& 6 & 6& 6 & & \textbf{2}\\ 
	\vdots&\vdots & \vdots& \vdots & \vdots & & \vdots \\
	29& 1& 5 & 5& 5 & & \textbf{2}\\ 
	30& 1& 6 & 6 & 6 & & 6\\         
\end{matrix}
\]
\caption{Sequences of events with the allocation of the receivers. All senders are fixed to Actor 1. Receivers in Sequence 1 are distributed equally, in Sequence 29 all the receivers are Actor 2.}
\label{seq}
\end{table}
By construction, in these relational event sequences the effect size of $\beta$ is smallest for the Sequence 1, grows gradually, and reaches its maximum for the Sequence 29. This data structure allows a clear comparison of the different shrinkage priors. The objective of this example is to check the amount and the shape of the shrinkage that the ridge, lasso, and horseshoe priors impose on a network effect. To see the clearest behaviors, we fix the shrinkage parameter $\lambda^2$ to 1 for all models.

\begin{figure}
	\centering
	\begin{subfigure}[b]{0.45\textwidth}
		\includegraphics[width=\textwidth]{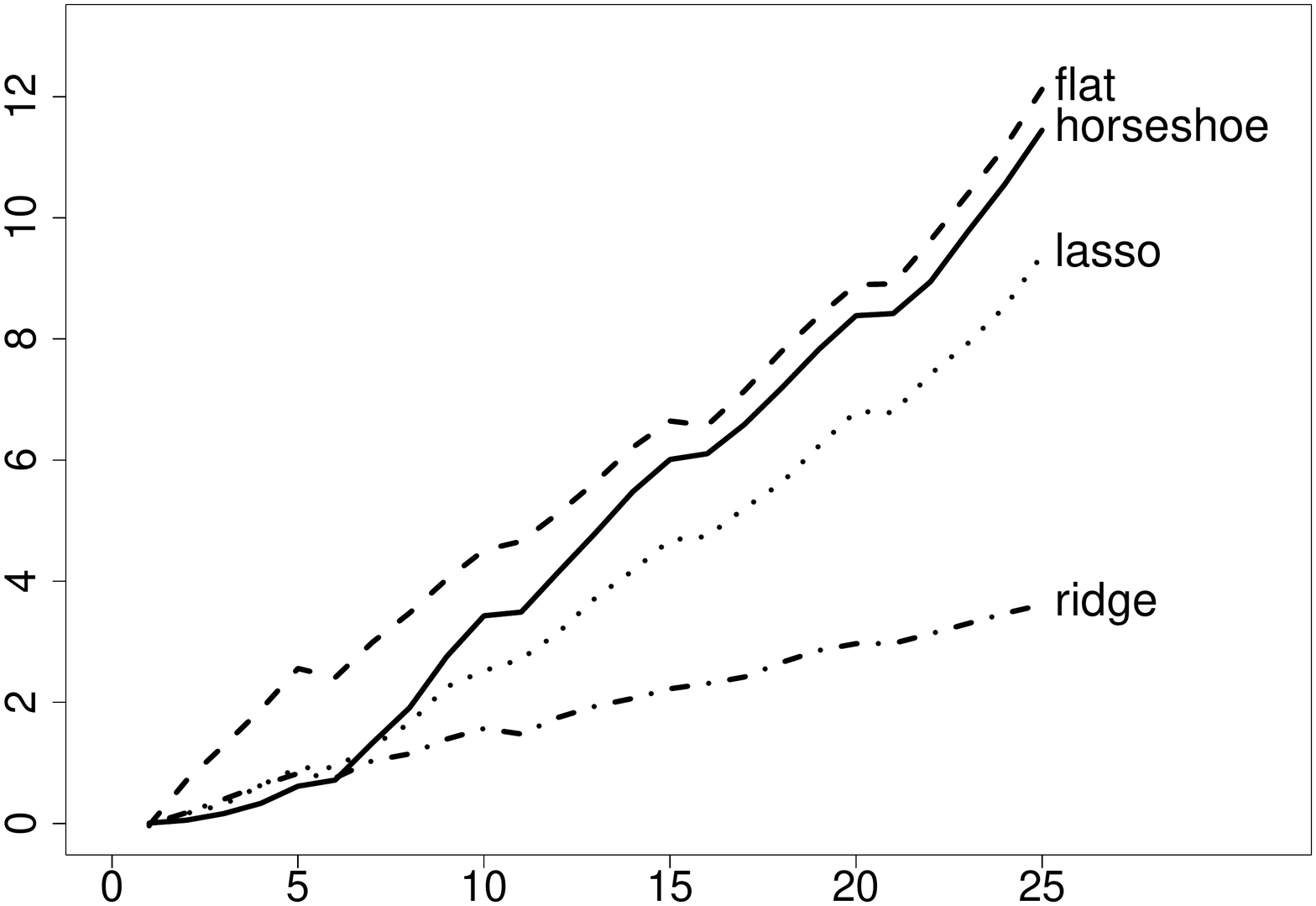}

		\caption{Estimated coefficient for the flat, ridge, lasso, and horseshoe models}
	\end{subfigure}
	\begin{subfigure}[b]{0.45\textwidth}
		\includegraphics[width=\textwidth]{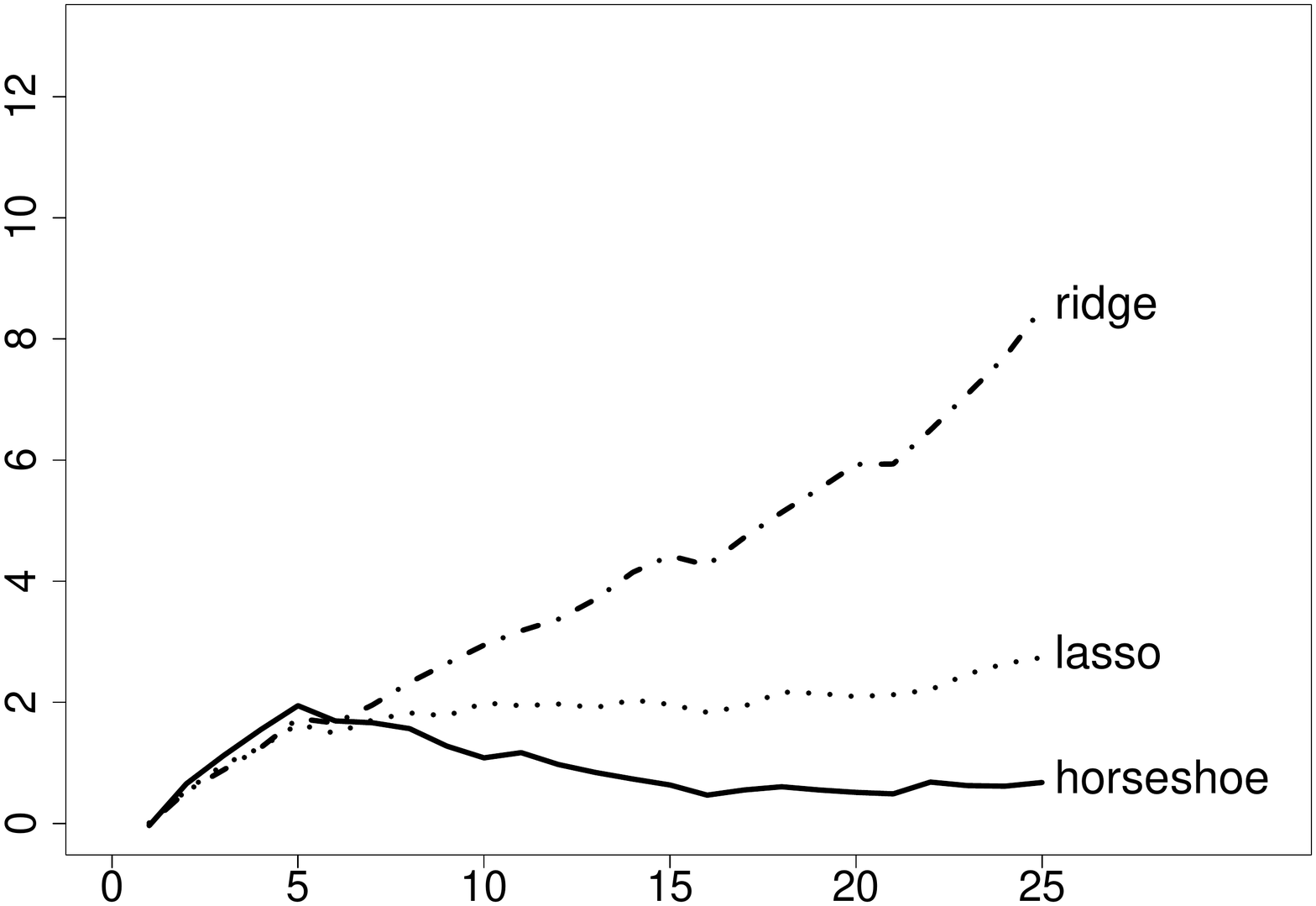}
		\caption{Estimated difference between the shrinkage models and the flat model}
	\end{subfigure}
	
	\caption{Estimated $\hat{\beta}$ for each generated sequence for the flat model (dashed line), ridge (dash-dotted line), lasso (dotted line), and horseshoe(solid line)}
	\label{fig:fixed_lambda}
\end{figure}

We used the actor oriented models to estimate the network effect for each of the 29 relational event sequences and show the resulting posterior means in Figure \ref{fig:fixed_lambda}. The left panel shows the estimate $\hat{\beta}$ across Sequences 1 to 29 and the right panel shows the difference between the estimate under the flat prior with the estimate under each shrinkage model. The estimate based on the flat prior model, which serves as a reference, correctly increases over the sequence index. The ridge model shows an approximate linear trend, heavily shrinking moderate to large effects. The lasso model shrinks small effects a bit more than the ridge prior and yields estimates that are approximately parallel those  of the flat model . The horseshoe shows most shrinkage for small effect sizes and the least shrinkage for larger effects where it gradually converges to the flat prior model. Next, we explore how the shrinkage prior models perform in empirical settings.

\section{Empirical data applications}

In this section we apply the Bayesian regularization algorithms to two empirical datasets. First we consider a relational event sequence based on the Enron corporate email data: the choice of the next receiver in this relational event sequence is analyzed using the actor-oriented models with shrinkage priors. Second, we consider a sequence of the communication messages from the famous Apollo 13 mission to the moon. This data is analyzed using the dyadic relational event models, in which the next dyad is predicted. The estimation is performed via R code which is available at  (github page link suppressed for blind review). The purpose of these empirical examples is to show that the shrinkage models (i) result in fewer significant coefficients compared to the flat no shrinkage prior model, and thus provide a more parsimonious description of the interaction behavior in the networks, and (ii) have comparable, and in some cases improved predictive power.

The evaluation of the model performance is assessed via the posterior predictive distribution (\cite{gelman2013bayesian}). In contrast to parametric bootstrapping (which only uses the MLE to predict observations), posterior predictive checks make use of the complete posterior of the parameters and thus take the uncertainty of the model parameters into account as well. Using the posterior predictive distribution, we can evaluate the performance of the fitted model by comparing the events predicted by the model with the events that actually occurred. Based on the posterior distribution, we can construct credible intervals and evaluate the significance of the particular effect by checking whether that interval covers zero. Additionally, to obtain point estimates we can look at the posterior mode of the effects based on the  sample from conditional posterior densities. 

To evaluate the predictive power of each model we use the latent variables $Z_i^{draw} = X_i\bm\beta^{draw}$ that are calculated on the sampled values $\bm\beta^{draw}$ generated by the Gibbs sampler. Each component of the vector $Z_i^{draw}$ corresponds to an actor from the ordered risk set. Higher values imply a larger probability of a corresponding actor to be the next receiver (or the dyad to transpire next in the case of dyad oriented model). Thus, we can compare the observed event with the set of ``best predicted events'', i.e. sorted values of latent variable $Z_i^{draw}$, for any event in the sequence $e_i, i \in 1, \dots, M$. Below, we consider how often events with the highest $Z_i^{draw}$ component are strictly equal to the observed event and how often it is in the set of 5\%, 10\%, or 20\% of the highest scored events. This measure can be calculated for both in-sample or out-of-sample predictions. 

\subsection{Analyzing email data using the actor-oriented model}

To demonstrate the performance of the actor oriented shrinkage models we used a publicly available data of the Enron corpus from the repository of Carnegie Mellon University. This dataset contains the time-stamped emails of 156 users, mostly senior management of Enron Corporation, a former American energy, commodities, and services company. The data were made public in 2001 after Enron Corporation declared bankruptcy and the following public investigation. It serves as a real life example of human communication, has been widely used in different fields, from social network research to computer science \citep{keila2005structure,  diesner2005communication, zhou2007strategies, 4983357, peterson2011email}, and has already been analyzed in the context of relational events in \cite{perry2013point} using maximum likelihood estimation with no shrinkage. 

Exogenous actor traits available in the data describe the actors' gender, department, seniority, job title, and company's division. We include interaction variables as $X(i)*Y(i)$, where $X$ and $Y$ come from the set of dichotomous actor dependent attributes $(L, T, J, F)$ -- see Table \ref{enron_actor_traits} for the overview. We also included dichotomous statistics describing whether sender and receiver have the same job title (1= yes, 0 = no)) whether they belong to the same division (1= yes, 0 = no). 

\begin{table}
\centering
\begin{tabular}{ll}
\hline 
Variable & Characteristics of actor i    \\
\hline 
L(i)     & member of the Legal department    \\
T(i)     & member of the Trading department  \\
J(i)     & seniority is Junior               \\
F(i)     & gender is Female                 \\
\hline
\end{tabular}
\caption{Dichotomous variables indicating whether the actor works in the Legal department, trading department, is a junior, or is female.
}
\label{enron_actor_traits}
\end{table}

As endogenous network effects, we include inertia and reciprocity -- statistics that are commonly included in the analysis of relational event data (\cite{leenders2016once}). The inertia statistic of the event $e_m = (t_m, s_m, r_m)$ measures the number of events sent from actor $s_m$ to actor $r_m$ until time $t_m$. The reciprocity statistic measures the number of events sent from actor $r_m$ to actor $s_m$ until time $t_m$. \cite{perry2013point} used interval-measured network effects to measure time dependence (see also \cite{Arena2021} for a memory decay models for relational event data). In this paper we adjusted time intervals to fit the chosen subset of events and calculated inertia and reciprocity on the intervals of 30 minutes, 1 hour, 2 hours, and 8 hours. The full list of effects used in the models is shown on Table \ref{enron_list}. 

\begin{table}[]
\footnotesize
\centering
\begin{tabular}{ll}
\hline
Effect          & Description                                              \\
\hline
\hline
\multicolumn{2}{c}{Endogenous dyadic effects}   \\
\hline 
Inertia 1       & Inertia on the interval of  less   than 30 minutes       \\
Inertia 2       & Inertia on the interval from 30 minutes to 1.1 hours     \\
Inertia 3       & Inertia on the interval from 1.1 hours to 2 hours        \\
Inertia 4       & Inertia on the interval from 2 hours to 8 hours          \\
Reciprocity 1   & Reciprocity on the interval of    less than 30 minutes   \\
Reciprocity 2   & Reciprocity on the interval from 30 minutes to 1.1 hours \\
Reciprocity 3   & Reciprocity on the interval from 1.1 hours to 2 hours    \\
Reciprocity 4   & Reciprocity on the interval from 2 hours to 8 hours      \\
\hline
\multicolumn{2}{c}{Exogenous dyadic effects}   \\
\hline 
Ex1  = L(i)*L(j) & Sender is from Legal department, receiver is from Legal department \\
Ex2  = T(i)*L(j) & Sender is from Trading department, receiver is from Legal department \\
Ex3  = J(i)*L(j) & Sender's title is Junior, receiver is from Legal department \\
Ex4  = F(i)*L(j) & Sender is female, receiver is from Legal department \\
Ex5  = L(i)*T(j) & Sender is from Legal department, receiver is from Trading department \\
Ex6  = T(i)*T(j) & Sender is from Trading department, receiver is from Trading department  \\
Ex7  = J(i)*T(j) & Sender's title is Junior, receiver is from Trading department \\
Ex8  = F(i)*T(j) & Sender is female, receiver is from Trading department \\
Ex9  = L(i)*J(j) & Sender is from Legal department, receiver's title is Junior \\
Ex10 = T(i)*J(j) & Sender is from Trading department, receiver's title is Junior \\
Ex11 = J(i)*J(j) & Sender's title is Junior, receiver's title is Junior \\
Ex12 = F(i)*J(j) & Sender is female, receiver's title is Junior \\
Ex13 = L(i)*F(j) & Sender is from Legal department, receiver is female \\
Ex14 = T(i)*F(j) & Sender is from Trading department, receiver is female \\
Ex15 = J(i)*F(j) & Sender's title is Junior, receiver is female \\
Ex16 = F(i)*F(j) & Sender is female, receiver is female \\
Same title       & Sender and receiver has the same title \\
Same   division  & Sender and receiver are from the same company division\\
\hline
\end{tabular}
\caption{List of effects for Enron Corpus data analysis, description of the exogenous events correspond to the covariate's value equal to one, and zero otherwise.}
\label{enron_list}
\end{table}

\begin{figure}
	\centering
	\begin{subfigure}[b]{0.9\textwidth}
		\includegraphics[width=\textwidth]{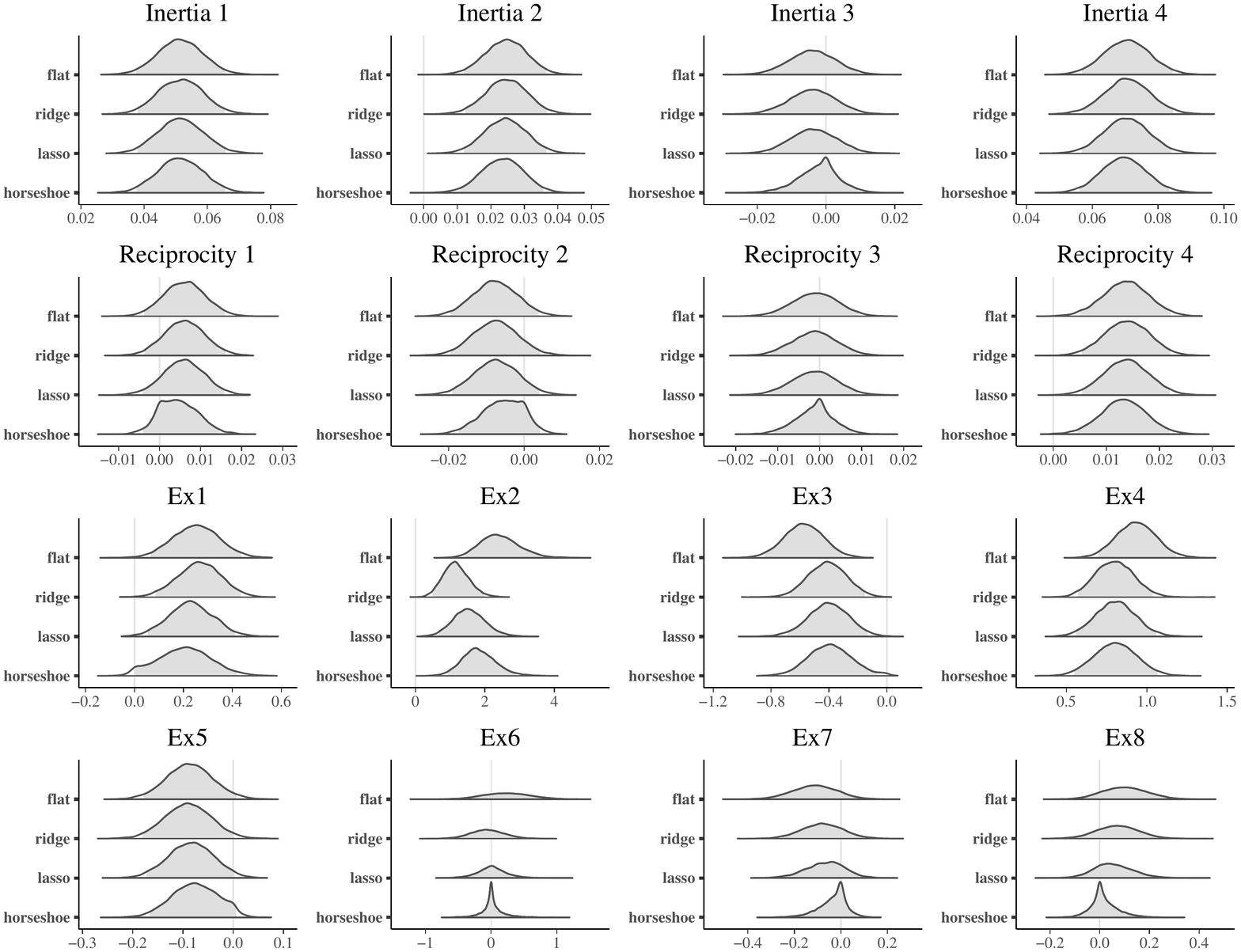}
	\end{subfigure}
	\begin{subfigure}[b]{0.9\textwidth}
		\includegraphics[width=\textwidth]{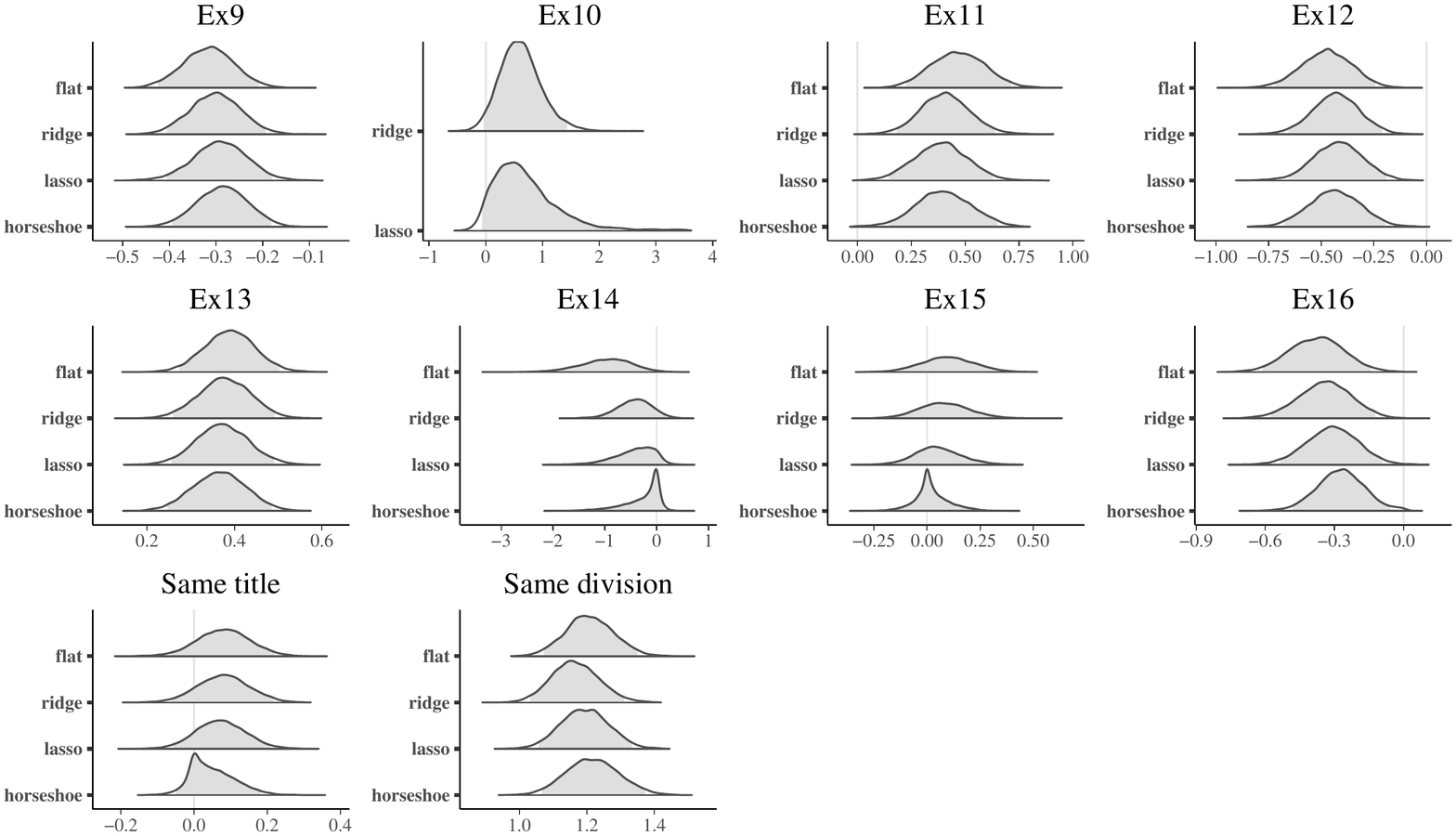}
	\end{subfigure}
	
	\caption{Density plots computed from posterior draws for Enron data estimation with 95\% credible intervals shown as shaded areas under the curves.}
	\label{fig:enron_ci}
\end{figure}

We estimated the four models on a subset of 1000 first events\footnote{Relational event models make the implicit assumption that effects are constant over the included event sequence. If effects change during the event sequence (e.g., when a new project is started or when a vacation period starts), the parameter estimates no longer meaningfully capture the interaction reality and predictive accuracy of the relational event model may decrease strongly \citep{mulder2019modeling}. Hence, we consider an event sequence that is short enough to reasonably assume stability of effects.} using the Gibbs sampler algorithms described in Section 2 with 10000 iterations as a burn-in period followed by a total of 100000 iterations, where only every tenth iteration is recorded (due to eliminate the effect of autocorrelation in the posterior draws). The estimated posterior distributions and 95\% credible intervals are plotted in Figure \ref{fig:enron_ci}. The credible intervals for the shrinkage models appear to be closer to zero, and the horseshoe often has a peak at zero. As expected, the flat prior model with no shrinkage returns the most significant predictors (fifteen); the ridge and the lasso model suggest fourteen significant predictors, and the horseshoe model results in the most parsimonious result with thirteen significant predictors (see Table \ref{enron_tableofci}).

Finally, we note that effects may not be identifiable for the flat prior model and the horseshoe model if there is not enough information in the data. In the analysis of the Enron data this problem occurred for 'Ex10', where the trajectory of the traceplot diverges. For the flat prior this can be caused by the prior being too `vague" and the information in the effect not being enough for the posterior to detect. Similarly, the prior density of the horseshoe is less informative than for the ridge and lasso models, due to its thicker tails. To avoid this identification problem for effect `Ex10', we excluded it from the estimation for the flat and horseshoe models. 

\begin{figure}[H]
	\centering
	\begin{subfigure}{0.49\textwidth}
		\includegraphics[width=\textwidth]{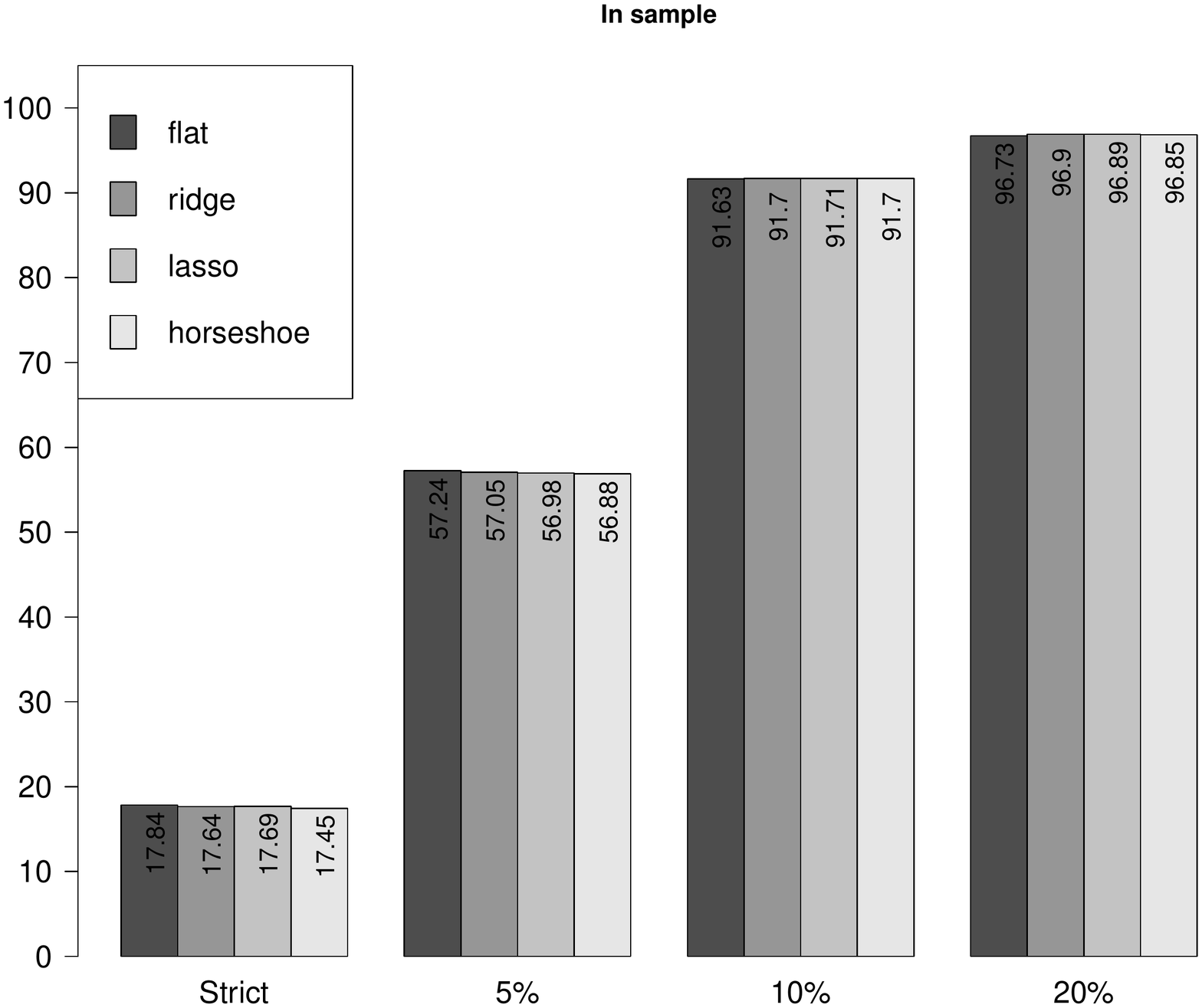}
		\caption{In-sample prediction}
		\label{enron_in}
	\end{subfigure}
	\begin{subfigure}{0.49\textwidth}
		\includegraphics[width=\textwidth]{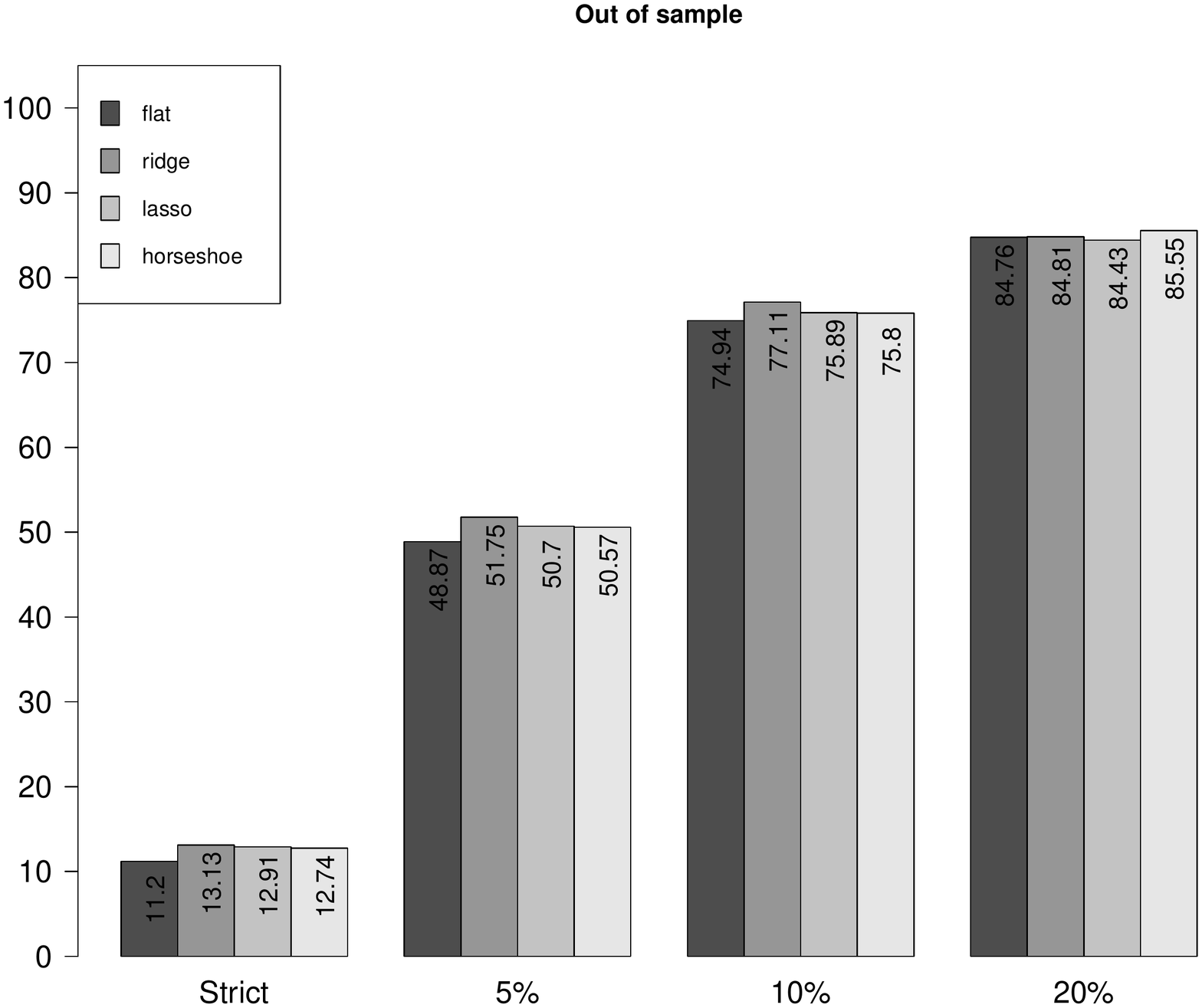}
		\caption{Out-of-sample prediction}
		\label{enron_out}
	\end{subfigure}
	
	\caption{Prediction score measures based on posterior draws, in percentages, for the Enron data}
	\label{fig:enron_prediction}
\end{figure}

Figure \ref{fig:enron_prediction} reports the values of the posterior predictive measure for in sample prediction(\ref{enron_in}) and out of sample prediction (\ref{enron_out}) calculated for the next 1000 events in the sequence. In the case of in-sample prediction, when we count the number of predicted events exactly equal to the observed events, we find exactly predicted interations in around 17\% of the events, with the flat model having the highest score of 17.84\%. In the 5\% range, the flat model still gives the best performance (the observed interaction is among the top 5\% predicted by the model in 57.2\% of the events-- while ridge, lasso and horseshoe have 57.1\%, 57.0\%, and 56.9\% respectively. In both the top 10\% and top 20\%, the shrinkage models perform (marginally) better than the no-shrinkage model. When considering out-of-sample prediction, where predictions were made for the next 1000 events, at least one shrinkage method shows higher scores than flat model in any score type.

\subsubsection{Extended number of potential effects}

\begin{figure}
	\centering
	\includegraphics[width = 0.8\textwidth]{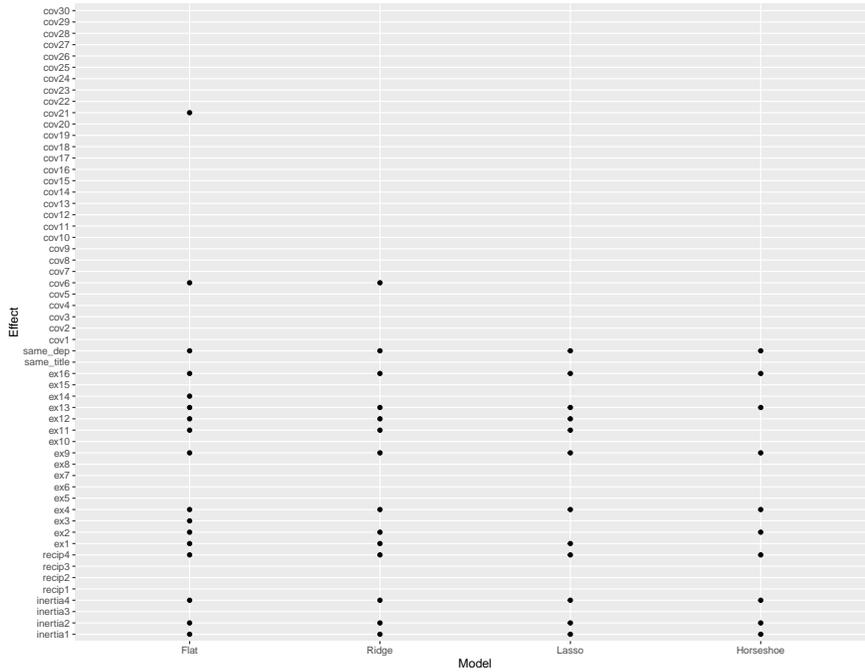}
	\caption{The resulting number of significant covariates based on credible intervals for Enron data}
	\label{enron_ext_dotplot}
\end{figure}

As a further test of the performance of the shrinkage models, we run a simple simulation procedure to assess the Type I error rate. The aim of a shrinkage model is to separate the wheat from the chaff: statistics that do not have an effect on the observed event sequence should be recognized as not significant by these models. In this manner, the effects that are marked as significant are then indeed likely to really play a role in the social process that caused these events to occur. We can not know which of the statistics included in the model above truly did (not) affect the actual event sequence. Therefore, we could only check whether the shrinkage models would deliver predictive performance that is comparable to the non-shrinkage model, while potentially using fewer statistics. We now test whether the models can recognize statistics for which we know their true effect is zero. We explore the ability of the models to recognize such immaterial variables by generating 30 random covariates. Since we generated these ourselves, we know for a fact that they could not have caused any of the events in the event sequence. Hence, ideally, none of them should end up being significant in any model. We add these 30 fake covariates to the model and run the analyses again. The results are shown in Figure \ref{enron_ext_dotplot}.

\begin{figure}
	\centering
	\begin{subfigure}{0.49\textwidth}
		\includegraphics[width=\textwidth]{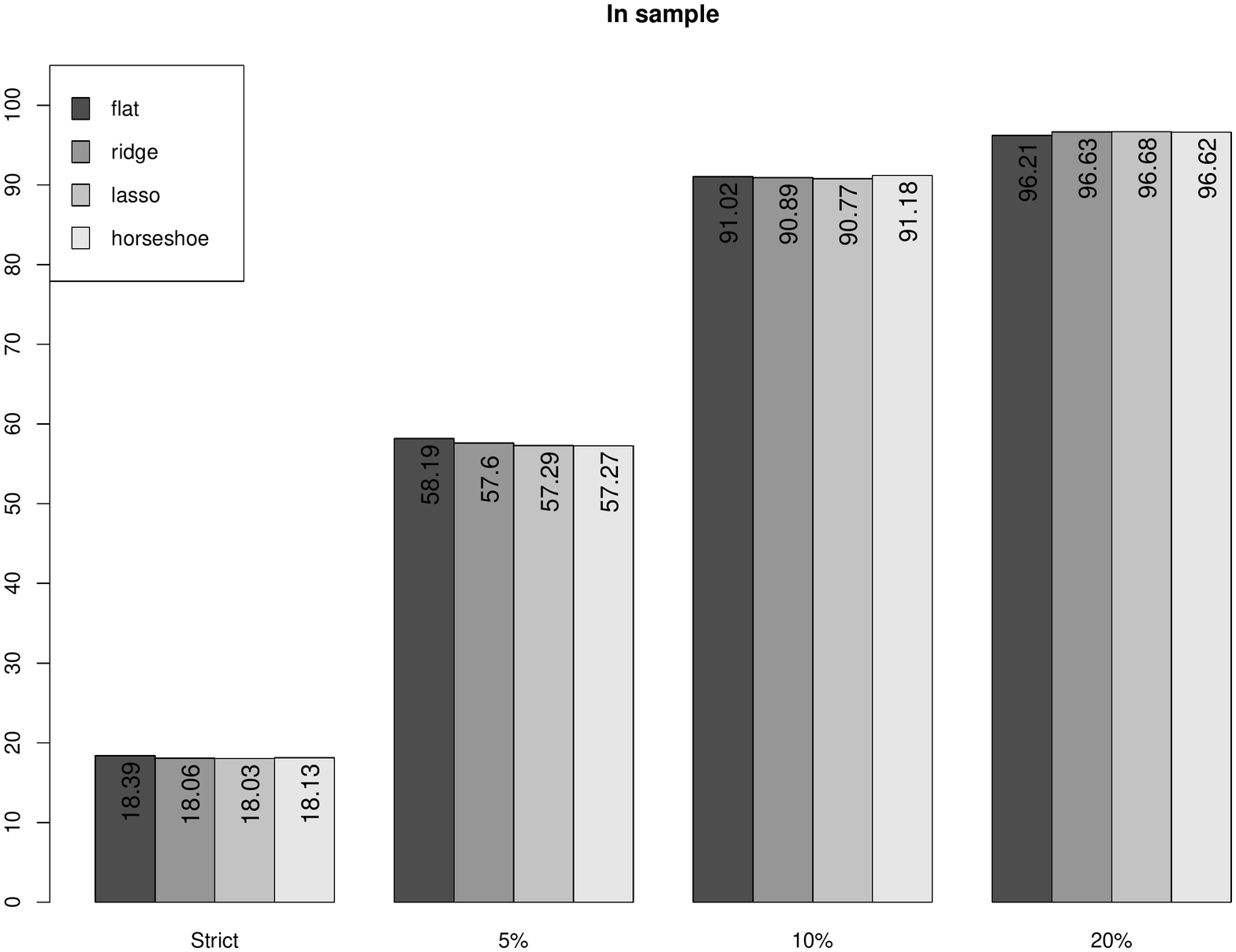}
		\caption{In-sample prediction}
		\label{enron_in}
	\end{subfigure}
	\begin{subfigure}{0.49\textwidth}
		\includegraphics[width=\textwidth]{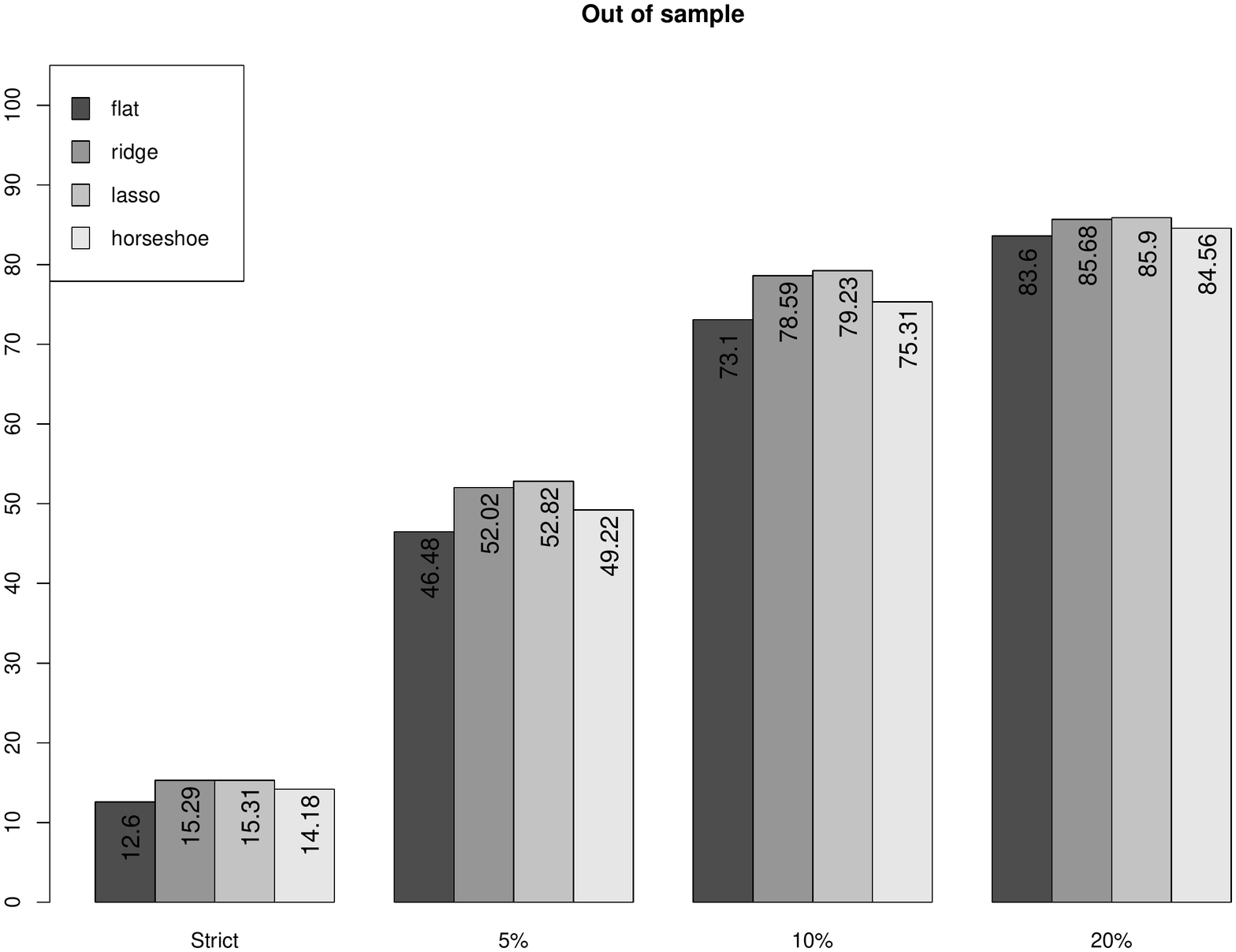}
		\caption{Out-of-sample prediction}
		\label{enron_out}
	\end{subfigure}
	
	\caption{Prediction score measures for the Enron data based on posterior draws, in percent (including the fake covariates)}
	\label{fig:enron_prediction_ext}
\end{figure}

The flat prior model with no shrinkage effect marks 17 of the 55 effects (without `Ex10') as significant, from which two effects are fake  variables. For the ridge prior model, only one of the fake effects is significant, with the total number of fourteen significant covariates out of 56 considered. The lasso and horseshoe models eliminate all fake effects, which confirms the expected behavior of the shrinkage models, resulting in twelve and ten significant effects respectively. In addition, we see that the predictive performance score has increased (compared to the analysis without the fake covariates) as a result of overfitting by the flat prior model (see Figure \ref{fig:enron_prediction_ext}) which is prevented by the shrinkage models. The shrinkage models (especially the lasso and horseshoe) do very well and separate the wheat from the chaff by shrinking the immaterial effects to zero, where the non-shrinkage model is fooled by some of them and reports two effects as significant of which we know for certain that they were unrelated to the observed event sequence. 

Furthermore, we notice that adding more effects to the model can inflate the overall shrinkage. For example, for the Enron data analysis lasso model detects fourteen predictors significant, while for the analysis with fake additional effects only twelve effects are significant. In the case of horseshoe model, these numbers are thirteen and ten respectively. More effects in the model can lead to larger sampled value of the shrinkage parameters, thus more effects are shrunk to zero.

\subsection{Analyzing the Apollo 13 data using the dyadic model}

	In this section we analyse the Flight Director voice loops from the infamous Apollo 13 mission. The data were retrieved from \url{http://apollo13realtime.org/} and consist of recorded voice messages between sixteen members of the team. In the original data only the senders of the messages were recorded, we therefore added the receivers manually (based on the actual text of the messages). The event sequence includes 4239 messages within the time interval of the first six hours following the Apollo 13 accident. As the mission underwent different periods (e.g., after and before ``Houston, we've had a problem'') the communication structure is likely to be different across these periods. This, again, suggests that the parameters of the relational event model likely change over time. Therefore we only consider the first 500 events which is a roughly stable period. For this data we estimate a dyad relational event model with 12 endogenous network effects: inertia, reciprocity, indegree, outdegree, and total degree for both sender and receiver, as well as participation shifts of the forms AB-BA, AB-BY, AB-XA, AB-AY (see \cite{butts20084} for the discussion of the network effects).

\begin{figure}
	\centering
	\includegraphics[width = 0.9\textwidth]{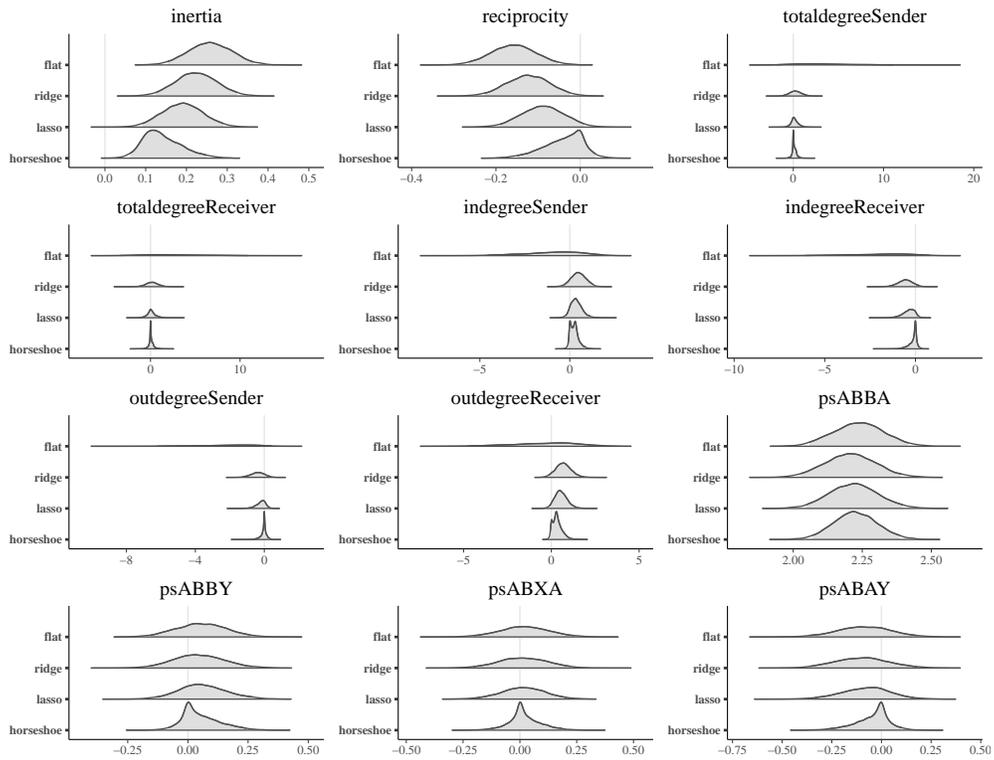}
	\caption{Posterior density plots and credible intervals for the Apollo 13 data.}
	\label{apollo_densities}
\end{figure}

We estimated the four models on a subset of 500 first events using the Gibbs sampler algorithms described in Section 2 with 10000 iterations in burn-in period and total 10000 iterations, where only every tenth iteration is stored. The estimated posteriors and the 95\% credible intervals are plotted in Figure \ref{apollo_densities}. The differences between the posteriors of the four models are similar to those in the analysis of the Enron data. For the reciprocity statistic the flat no-shrinkage model does not include zero in the credible interval (neither does the ridge regression model). Alternatively, credible intervals for reciprocity in the lasso and horseshoe model do include zero, and for the horseshoe the probability mass is clearly concentrated at zero. The ridge prior model and the flat prior model each result in three significant effects. The horseshoe and lasso shrinkage models on the other hand eliminate the small negligible effect of reciprocity, resulting in two significant effects. 

It is uncommon for a reciprocity effect to be zero in most social settings, especially in highly regulated situations as in a space flight. In fact, in our dataset, 259 events are sent and 234 are received by the flight director, mostly in back-and-forth interaction. This indicates an importance of the reciprocal behaviour for our dataset. Whereas the reciprocity statistic as defined by \cite{butts20084} captures a general trend towards interacting with past senders, the alternative statistic AB-BA captures \textit{immediate} reciprocity: after A sends a message to B, B reciprocates this immediately. AB-BA statistic describes better the instant reciprocal behaviour in this data and has a large effect of around 2.25 that stays intact in the shrinkage models.  In addition, we noticed that inertia   and reciprocity statistics have a correlation of 0.97. Even with the presence of strong collinearity among the predictors, the lasso and horseshoe shrinkage models are able to eliminate the effect of reciprocity from the analysis.

\begin{figure}
	\centering
	\begin{subfigure}{0.49\textwidth}
		\includegraphics[width=\textwidth]{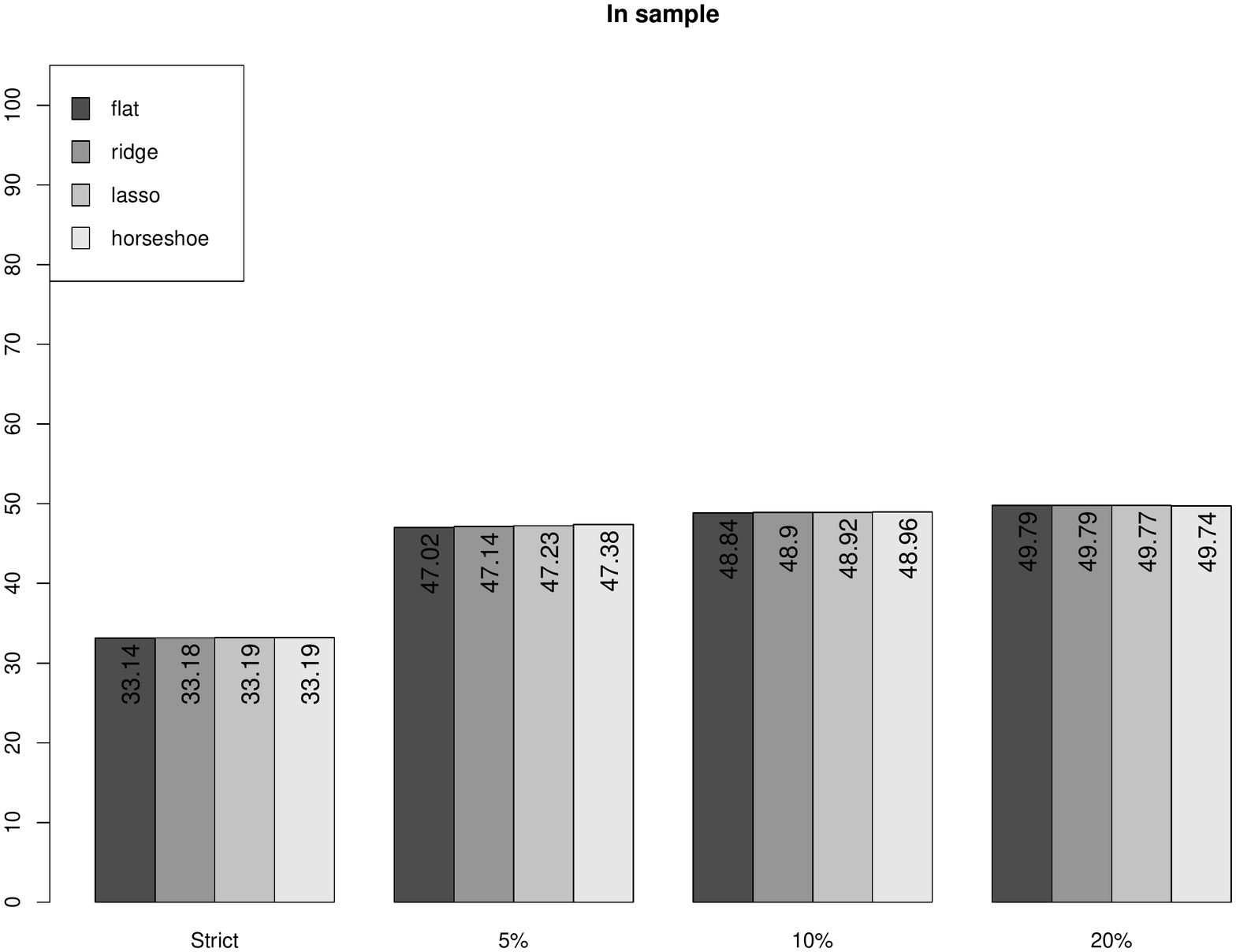}
		\caption{In-sample prediction}
		\label{apollo_in}
	\end{subfigure}
	\begin{subfigure}{0.49\textwidth}
		\includegraphics[width=\textwidth]{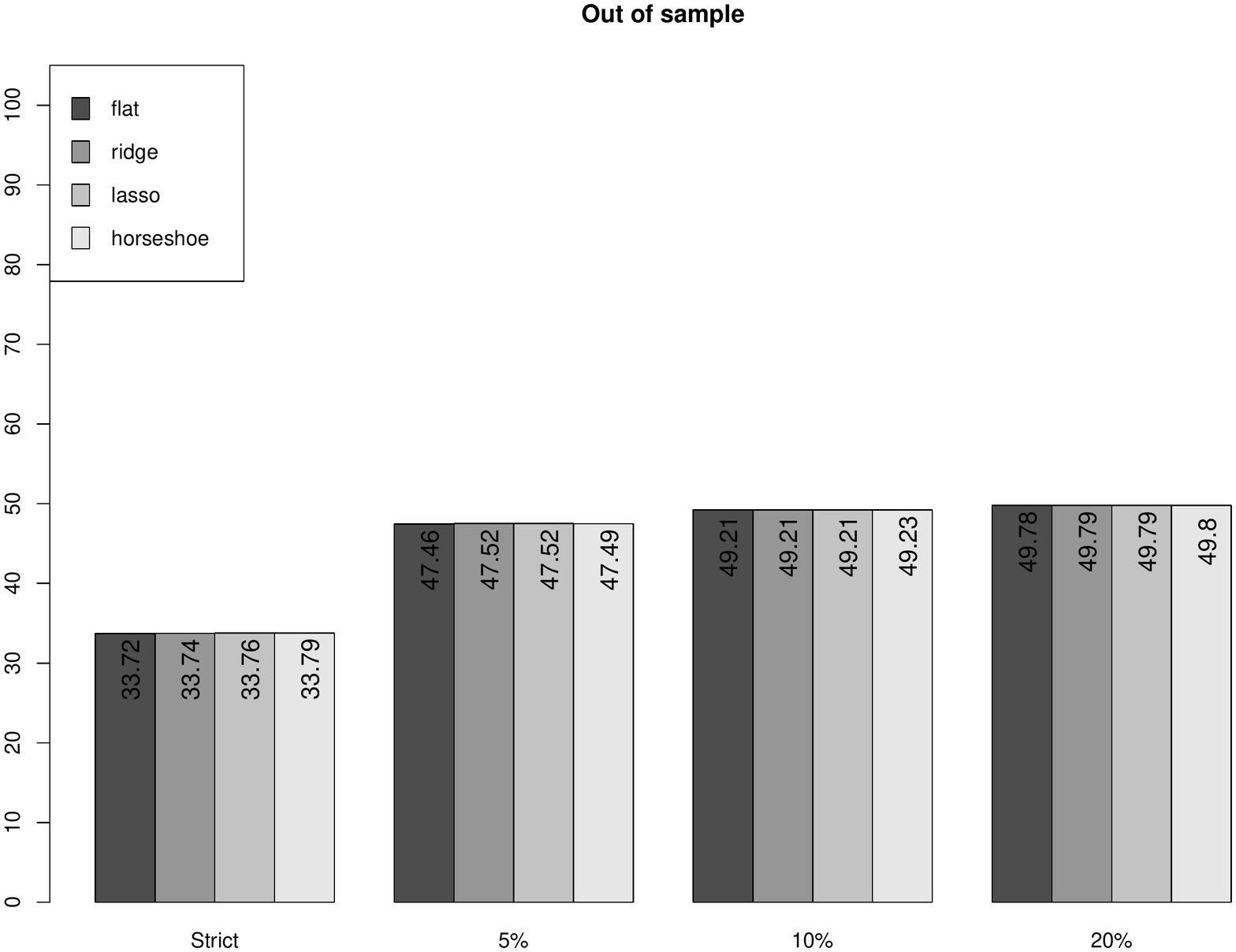}
		\caption{Out-of-sample prediction}
		\label{apollo_out}
	\end{subfigure}
	
	\caption{Prediction scores for the Apollo 13 data, in percent, based on posterior draws}
	\label{fig:apollo_prediction}
\end{figure}

Figure \ref{fig:apollo_prediction} reports the values of the posterior predictive measure for in sample prediction(\ref{apollo_in}) and out of sample prediction(\ref{apollo_out}). We see comparable predictive performance between the models, with marginal but constant improvement for the shrinkage methods compared to the flat prior model. The same observation holds for the out-of-sample prediction. The shrinkage models do this with more parsimonious and more easily interpreted models. While the shrinkage model eliminated a negligible effect of reciprocity statistic, the prediction score for shrinkage models does not decrease much compared to the one for flat model.

\subsubsection{Extended number of potential effects}

\begin{figure}
\centering
\includegraphics[width = 0.8\textwidth]{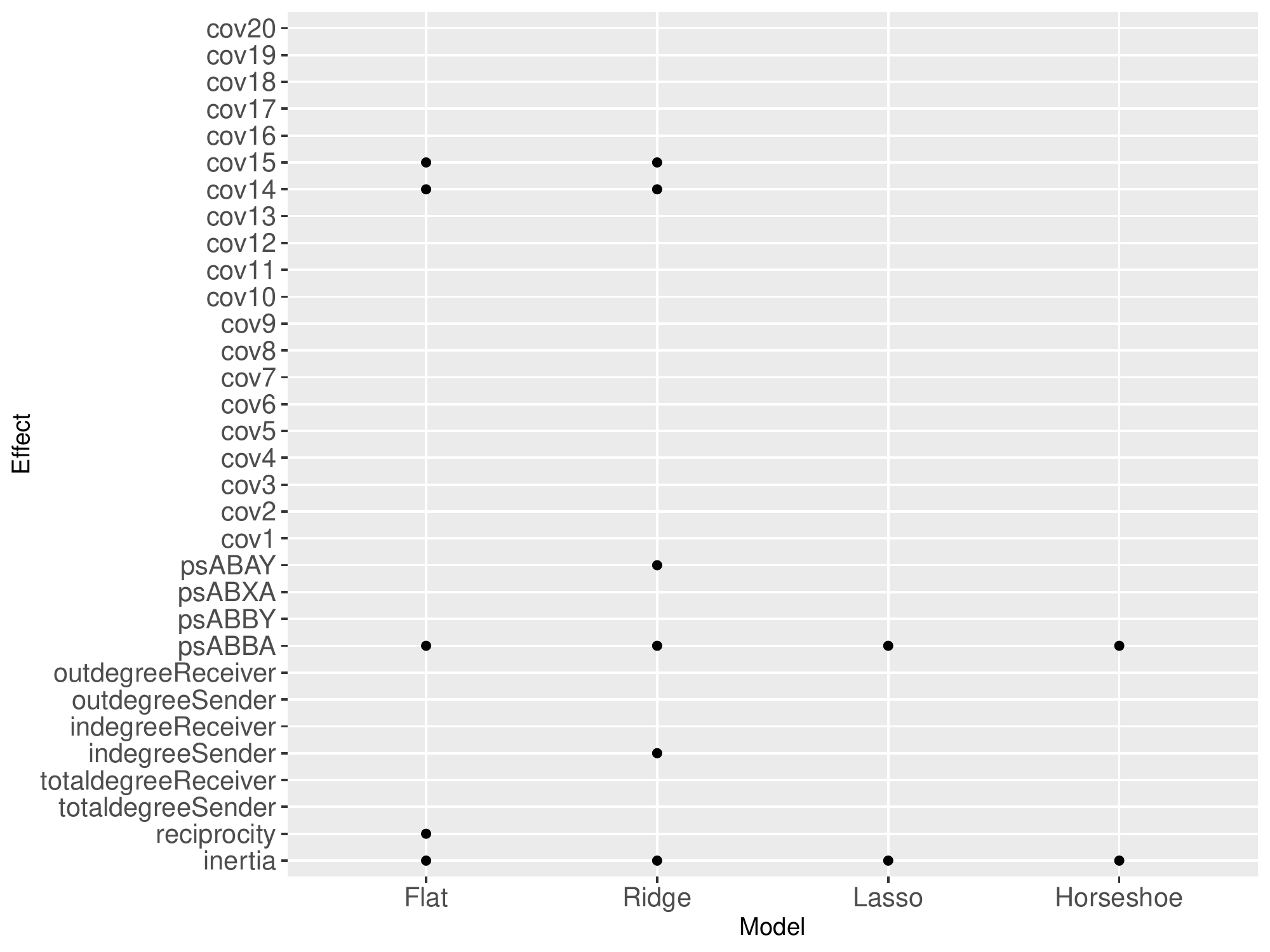}
\caption{The resulting number of significant covariates for the Apollo 13 data}
\label{apollo_ext_dotplot}
\end{figure}

The statistics that are found to be significant in the models make great substantive sense and would be expected to be found for an interaction sequence as a space flight. We now extend the dataset by adding randomly generated (fake) statistics to the model, similarly to the analysis of the Enron data in Section 4.1. Figure \ref{apollo_ext_dotplot} displays which effects are considered significant by each of the models, based on the 95\% credible intervals. The flat and ridge models mark two of the fake covariates as significant, while lasso and horseshoe models with stronger shrinkage effect are not tricked and do not mark any of the fake covariates as significant resulting in a parsimonious and more correct model. 

\begin{figure}[h]
	\centering
	\begin{subfigure}{0.49\textwidth}
		\includegraphics[width=\textwidth]{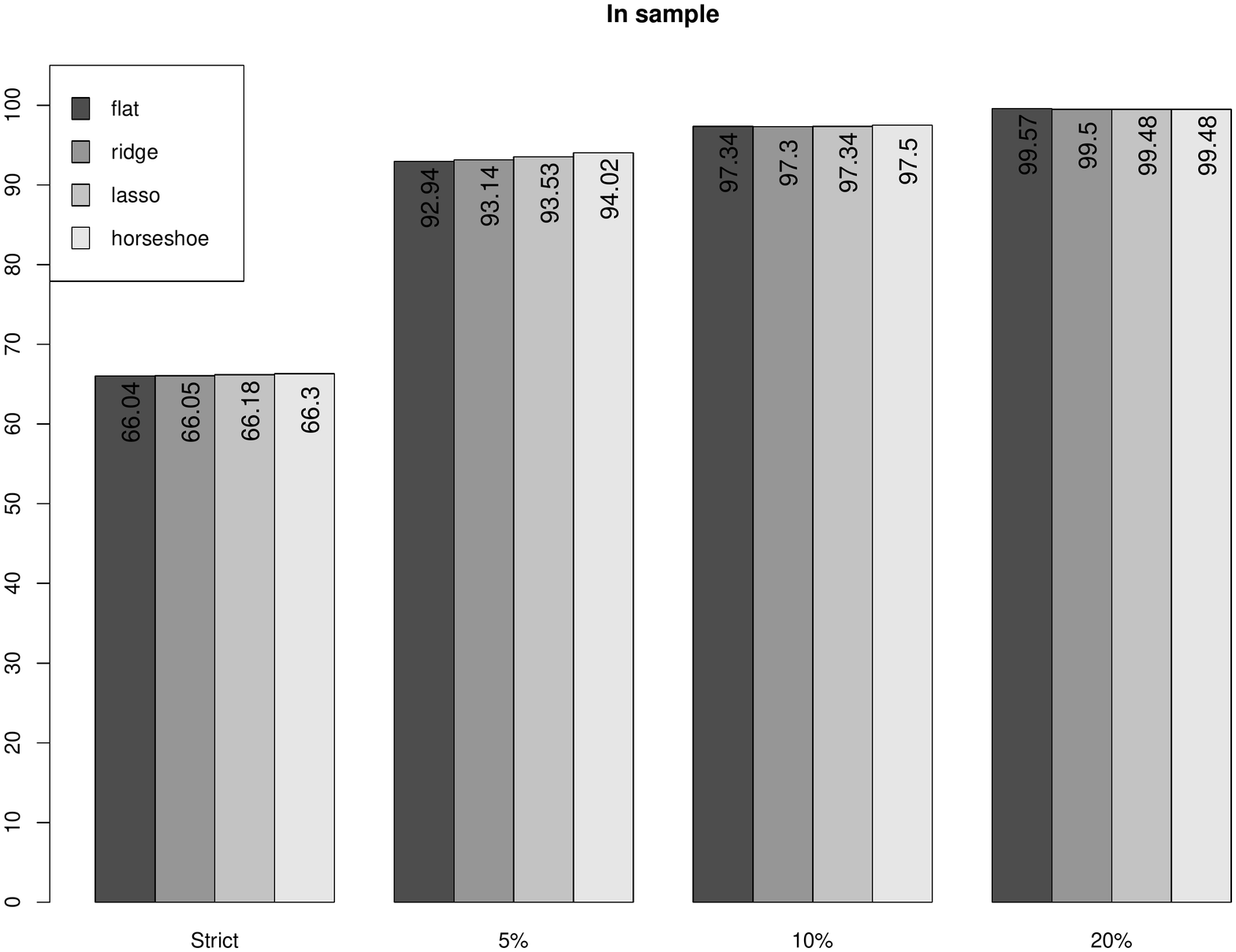}
		\caption{In-sample prediction}
		\label{apollo_ext_in}
	\end{subfigure}
	\begin{subfigure}{0.49\textwidth}
		\includegraphics[width=\textwidth]{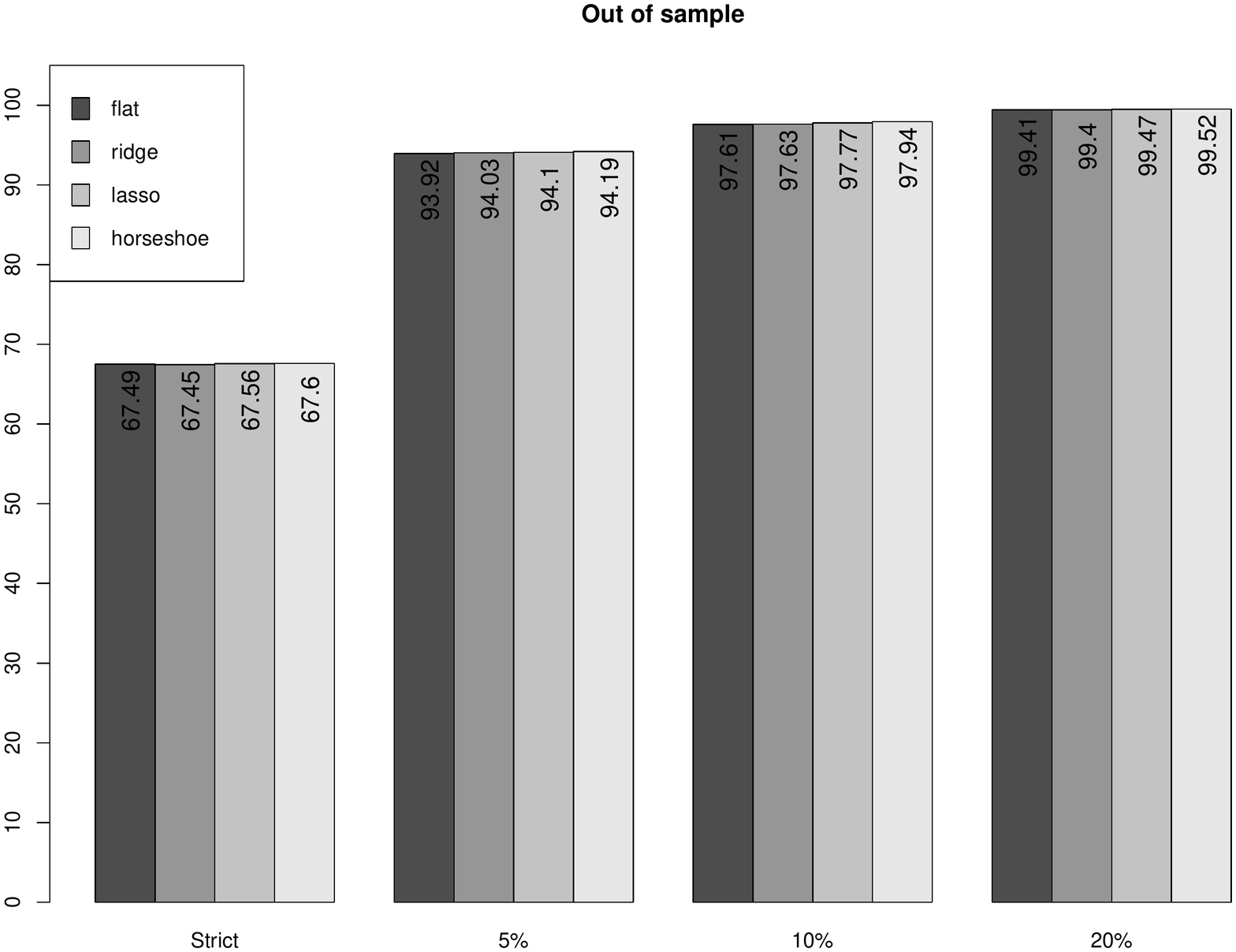}
		\caption{Out-of-sample prediction}
		\label{apollo_ext_out}
	\end{subfigure}
	
	\caption{Prediction score measures for the Apollo 13 data, in percent, based on posterior draws (including the fake covariates)}
	\label{fig:apollo_prediction_ext}
\end{figure}

Figure \ref{fig:apollo_prediction_ext} reports the values of the posterior predictive measures for in-sample prediction(\ref{apollo_ext_in}) and out-of-sample prediction(\ref{apollo_ext_out}) based on the models with the full data sets (including the fake covariates). The scores are considerably higher than for the primary analysis without the additional fake covariates; this supports the hypothesis that the shrinkage models do perform better in the case of sparse effects. In both in-sample and out-of-sample cases, the no shrinkage model has the lowest performance in almost in all cases. Among the shrinkage models, the horseshoe and lasso  models slightly outperform the ridge regression model. We conclude that the shrinkage models were able to recognize the fake statistics as unconnected to the data generation process. As a result, they are less prone to overfitting or to added noise induced by irrelevant predictors.

\section{Discussion}
The scope of this paper was to extend Bayesian variable selection methods to relational event data analyses. Recent research shows that the evolution of relational event networks and the formation of the links between actors can be triggered by a great variety of (complex) effects (both endogenous network effects and exogenous statistics, as well as interaction effects). Therefore, there is a need for statistical algorithms that are able to separate the wheat from the chaff to recognize the true nonzero effects. This results in the more parsimonious relational event models that are easier to interpret and have good predictive performance. The proposed Bayesian shrinkage models provide an effective solution to this problem.

The Bayesian shrinkage methods show promising results in detecting relevant predictors without sacrificing predictive accuracy (and in some cases even improving prediction accuracy). Furthermore, by considering a full Bayesian approach, the shrinkage parameter is jointly estimated with the other model parameters, and thus no (ad hoc) cross-validation methods are needed to determine the shrinkage parameters as would be the case in a non-Bayesian penalized regression. Moreover, we obtain a clear Bayesian interpretation of the results where parameter uncertainty is naturally incorporated -- this is not the case with the classical lasso which may, in turn, result in standard errors of zero. A further advantage of shrinkage priors is that they can avoid identification issues and guarantee convergence of the Gibbs sampler.

When  prior information about the possible distribution of effects is not available, our recommendation is to use all three models discussed in this paper and compare their posterior predictive measures afterwards. The horseshoe prior generally results in the most parsimonious model because it can have zeros as the point estimates (posterior modes). This tends to lead to the model that is the easiest to interpret with effects that are the least likely to be significant ``by accident". This makes the final modeling more conservative, which is probably a good thing considering that there is little (if any) solid theory to base the choice of predictors in relational event models on.

Our study suggests that adding a large number of predictors in the model may lead to a magnified shrinkage effect. This can be explained by the fact that the shrinkage parameter is optimized based on all the predictors in the model. Thus, if more variables that are added have an approximate zero effect, the average degree of shrinkage of all effects will be slightly larger. If, on the other hand, variables would be added with relatively large, nonzero effects, the average degree of shrinkage of all effects will be slightly smaller. Note that a similar behavior would be observed in classical regularization methods. As a possible remedy, one can define separate shrinkage priors for different sets of the effects in the model to control the average amount of shrinkage differently across these different sets. This would be an interesting extension for future work on Bayesian regularization of relational event models. A further study could assess the possible benefits of such fine-grained shrinkage methods.

In addition, further research can explore the use of shrinkage priors to model temporal changes of network parameters over time. The idea would be to place a shrinkage prior for the difference of a parameter over time, similar as in graphical models for variables \citep{shafiee2020non}. When the parameter is approximately constant over time, the difference is shrunk to zero such that the estimated parameter remains constant. If the parameter does change, e.g., due to a switch between interaction regimes between the actors, no shrinkage should be applied and we would be able to identify a change of the interaction regime in the network.

\pagebreak
\appendix
\section{Tables and graphs}
\begin{table}[H]
	\scriptsize
	\centering
	\begin{tabular}{lcccc}
		\multicolumn{1}{l}{}        & \multicolumn{4}{c}{Models}    \\ 
		\cline{2-5} 	
		\multicolumn{1}{c}{Effect}   & \multicolumn{1}{c}{Flat}  & \multicolumn{1}{c}{Ridge} & \multicolumn{1}{c}{Lasso} & \multicolumn{1}{c}{Horseshoe} \\
		\hline 
	Inertia 1     & (0.0374, 0.0655)   & (0.0379, 0.0655)   & (0.0378, 0.0657)   & (0.0375, 0.0656)   \\
	Inertia 2     & (0.012, 0.0365)    & (0.0124, 0.0368)   & (0.012, 0.0366)    & (0.0108, 0.0357)   \\
	Inertia 3     & (-0.0174, 0.0093)  & (-0.0173, 0.0089)  & (-0.0168, 0.0093)  & (-0.0152, 0.0093)  \\
	Inertia 4     & (0.0568, 0.0843)   & (0.057, 0.0846)    & (0.0568, 0.084)    & (0.056, 0.0834)    \\
	Reciprocity 1 & (-0.0038, 0.0159)  & (-0.0037, 0.0161)  & (-0.0042, 0.0158)  & (-0.004, 0.0145)   \\
	Reciprocity 2 & (-0.0195, 0.0038)  & (-0.0195, 0.0042)  & (-0.0191, 0.0041)  & (-0.0173, 0.0042)  \\
	Reciprocity 3 & (-0.0118, 0.0093)  & (-0.0118, 0.0091)  & (-0.0119, 0.009)   & (-0.0109, 0.0078)  \\
	Reciprocity 4 & (0.005, 0.0221)    & (0.0054, 0.0221)   & (0.0054, 0.022)    & (0.0046, 0.0219)   \\
	Ex1           & (0.0651, 0.4367)   & (0.0864, 0.4342)   & (0.0532, 0.4113)   & (-0.0003, 0.4175)  \\
	Ex2           & (1.4358, 3.5986)   & (0.491, 1.9644)    & (0.6754, 2.5357)   & (0.9374, 2.8983)   \\
	Ex3           & (-0.8426, -0.305)  & (-0.6689, -0.1522) & (-0.6878, -0.1317) & (-0.6775, -0.066)  \\
	Ex4           & (0.6988, 1.1809)   & (0.5721, 1.0373)   & (0.5611, 1.0591)   & (0.549, 1.0691)    \\
	Ex5           & (-0.1782, 0.0065)  & (-0.1806, 0.002)   & (-0.1727, 0.003)   & (-0.1659, 0.0059)  \\
	Ex6           & (-0.4597, 0.9358)  & (-0.5537, 0.4543)  & (-0.458, 0.504)    & (-0.3299, 0.4403)  \\
	Ex7           & (-0.3, 0.0733)     & (-0.2553, 0.0976)  & (-0.2432, 0.0894)  & (-0.1801, 0.0674)  \\
	Ex8           & (-0.0706, 0.2764)  & (-0.0986, 0.2387)  & (-0.09, 0.2231)    & (-0.0832, 0.1472)  \\
	Ex9           & (-0.4247, -0.2086) & (-0.4057, -0.1937) & (-0.4057, -0.1787) & (-0.3939, -0.1817) \\
	Ex10          &                    & (-0.0331, 1.4379)  & \multicolumn{2}{l}{(-0.0648, 2.2296)}   \\
	Ex11          & (0.2266, 0.714)    & (0.1825, 0.6399)   & (0.1455, 0.6281)   & (0.1454, 0.6501)   \\
	Ex12          & (-0.7198, -0.2434) & (-0.6555, -0.2116) & (-0.6547, -0.1912) & (-0.6751, -0.1962) \\
	Ex13          & (0.2737, 0.5034)   & (0.2596, 0.4921)   & (0.2556, 0.4913)   & (0.2463, 0.4853)   \\
	Ex14          & (-2.0284, -0.0974) & (-1.1362, 0.1878)  & (-1.3042, 0.1454)  & (-1.2347, 0.1396)  \\
	Ex15          & (-0.1411, 0.3399)  & (-0.1496, 0.3241)  & (-0.1534, 0.2769)  & (-0.1395, 0.2143)  \\
	Ex16          & (-0.6111, -0.1409) & (-0.576, -0.1227)  & (-0.5362, -0.0911) & (-0.4818, -0.0565) \\
	Same title    & (-0.0617, 0.2192)  & (-0.0593, 0.2171)  & (-0.0574, 0.2074)  & (-0.0488, 0.1832)  \\
	Same division & (1.0716, 1.3499)   & (1.0323, 1.3105)   & (1.0565, 1.3337)   & (1.0664, 1.3732)  \\
	\hline
	Number of significant covariates & 15 & 14 & 14& 13
	\end{tabular}
	\caption{95\% Credible intervals for the Enron corpus data.}
	\label{enron_tableofci}
\end{table}

\begin{table}[]
	\centering
\scriptsize
	\begin{tabular}{l|cccc|cccc}
		& \multicolumn{4}{c|}{First analysis}   & \multicolumn{4}{c}{Extended analysis} \\
		& Flat   & Ridge  & Lasso  & Horseshoe & Flat    & Ridge  & Lasso  & Horseshoe \\
		\hline
		\hline
	Inertia 1     & 0.05   & 0.052  & 0.052       & 0.05        & 0.055  & 0.053  & 0.053  & 0.052  \\
	Inertia 2     & 0.024  & 0.023  & 0.024       & 0.021       & 0.024  & 0.026  & 0.026  & 0.022  \\
	Inertia 3     & -0.003 & -0.003 & -0.002      & 0           & -0.004 & -0.004 & -0.005 & 0      \\
	Inertia 4     & 0.072  & 0.07   & 0.072       & 0.068       & 0.07   & 0.071  & 0.068  & 0.067  \\
	Reciprocity 1 & 0.007  & 0.006  & 0.006       & 0           & 0.006  & 0.005  & 0.005  & 0      \\
	Reciprocity 2 & -0.009 & -0.007 & -0.008      & 0           & -0.007 & -0.008 & -0.008 & 0      \\
	Reciprocity 3 & 0      & 0      & -0.002      & 0           & -0.001 & 0      & -0.001 & 0      \\
	Reciprocity 4 & 0.013  & 0.013  & 0.014       & 0.014       & 0.014  & 0.014  & 0.014  & 0.013  \\
	Ex1           & 0.248  & 0.256  & 0.219       & 0.205       & 0.264  & 0.348  & 0.223  & 0      \\
	Ex2           & 2.341  & 1.19   & 1.551       & 1.723       & 2.211  & 0.555  & 0.418  & 1.604  \\
	Ex3           & -0.605 & -0.433 & -0.42       & -0.403      & -0.564 & -0.213 & -0.113 & -0.001 \\
	Ex4           & 0.912  & 0.808  & 0.77        & 0.837       & 0.913  & 0.636  & 0.61   & 0.735  \\
	Ex5           & -0.09  & -0.094 & -0.09       & -0.076      & -0.054 & -0.037 & -0.032 & 0      \\
	Ex6           & 0.276  & -0.101 & 0.005       & 0           & 0.319  & -0.091 & -0.008 & -0.001 \\
	Ex7           & -0.104 & -0.095 & -0.044      & -0.001      & -0.125 & -0.032 & -0.024 & 0      \\
	Ex8           & 0.113  & 0.074  & 0.029       & 0           & 0.171  & 0.055  & -0.001 & 0      \\
	Ex9           & -0.313 & -0.297 & -0.286      & -0.274      & -0.323 & -0.293 & -0.259 & -0.275 \\
	Ex10          &        & 0.602  & \multicolumn{2}{l}{0.482} &        & 0.27   & 0.008  &        \\
	Ex11          & 0.449  & 0.398  & 0.418       & 0.385       & 0.456  & 0.358  & 0.257  & -0.001 \\
	Ex12          & -0.468 & -0.422 & -0.388      & -0.409      & -0.513 & -0.381 & -0.355 & 0.001  \\
	Ex13          & 0.392  & 0.379  & 0.373       & 0.356       & 0.393  & 0.356  & 0.356  & 0.346  \\
	Ex14          & -0.805 & -0.416 & -0.168      & -0.002      & -0.896 & -0.094 & -0.008 & 0      \\
	Ex15          & 0.143  & 0.044  & 0.025       & 0.001       & 0.113  & 0.035  & 0.003  & 0      \\
	Ex16          & -0.356 & -0.314 & -0.319      & -0.284      & -0.382 & -0.246 & -0.213 & -0.273 \\
	Same title    & 0.09   & 0.086  & 0.088       & 0           & 0.077  & 0.077  & 0.06   & 0      \\
	Same division & 1.169  & 1.163  & 1.22        & 1.189       & 1.194  & 1.099  & 1.178  & 1.295  \\
	\hline
	Cov1          &        &        &             &             & 0.011  & 0.013  & 0.01   & 0      \\
	Cov2          &        &        &             &             & -0.026 & -0.026 & -0.019 & 0      \\
	Cov3          &        &        &             &             & 0.01   & 0.011  & 0.008  & 0      \\
	Cov4          &        &        &             &             & -0.025 & -0.019 & -0.01  & 0      \\
	Cov5          &        &        &             &             & 0.003  & 0.004  & 0.008  & 0      \\
	Cov6          &        &        &             &             & -0.048 & -0.052 & -0.033 & 0      \\
	Cov7          &        &        &             &             & -0.007 & -0.004 & -0.002 & 0      \\
	Cov8          &        &        &             &             & -0.02  & -0.044 & -0.016 & 0      \\
	Cov9          &        &        &             &             & -0.047 & -0.065 & -0.041 & 0      \\
	Cov10         &        &        &             &             & -0.004 & 0      & 0.003  & 0      \\
	Cov11         &        &        &             &             & 0      & -0.001 & 0      & 0      \\
	Cov12         &        &        &             &             & -0.008 & -0.007 & -0.002 & 0      \\
	Cov13         &        &        &             &             & -0.02  & -0.028 & -0.026 & 0      \\
	Cov14         &        &        &             &             & -0.003 & -0.004 & -0.003 & 0      \\
	Cov15         &        &        &             &             & -0.005 & -0.007 & -0.002 & 0      \\
	Cov16         &        &        &             &             & 0.018  & 0.016  & 0.008  & 0      \\
	Cov17         &        &        &             &             & -0.001 & 0      & -0.001 & 0      \\
	Cov18         &        &        &             &             & 0.002  & -0.005 & -0.002 & 0      \\
	Cov19         &        &        &             &             & -0.005 & -0.001 & 0      & 0      \\
	Cov20         &        &        &             &             & -0.057 & -0.059 & -0.049 & 0      \\
	Cov21         &        &        &             &             & -0.056 & -0.05  & -0.045 & 0      \\
	Cov22         &        &        &             &             & -0.012 & -0.007 & -0.001 & 0      \\
	Cov23         &        &        &             &             & 0.052  & 0.038  & 0.034  & 0      \\
	Cov24         &        &        &             &             & -0.038 & -0.025 & -0.024 & 0      \\
	Cov25         &        &        &             &             & -0.017 & -0.009 & -0.019 & 0      \\
	Cov26         &        &        &             &             & -0.004 & -0.002 & 0.005  & 0      \\
	Cov27         &        &        &             &             & -0.025 & -0.022 & -0.018 & 0      \\
	Cov28         &        &        &             &             & -0.029 & -0.034 & -0.024 & 0      \\
	Cov29         &        &        &             &             & -0.023 & -0.02  & -0.012 & 0      \\
	Cov30         &        &        &             &             & -0.001 & 0.009  & -0.006 & 0     \\
		\hline
		\hline        
	\end{tabular}
\label{enron_mode}
\caption{Values of the posterior mode, for effects in primary and extended model of the Enron data.}
\end{table}

\begin{table}[]
	\scriptsize
	\begin{tabular}{lcccc}
		\multicolumn{1}{l}{} & \multicolumn{4}{c}{Models}    \\ 
	\cline{2-5} 	
	\multicolumn{1}{c}{Effect}   & \multicolumn{1}{c}{Flat}  & \multicolumn{1}{c}{Ridge} & \multicolumn{1}{c}{Lasso} & \multicolumn{1}{c}{Horseshoe} \\
	\hline 
		inertia                          & (0.1534, 0.3637)   & (0.1252, 0.3245)   & (0.092, 0.2903)   & (0.0619, 0.2493)  \\
		reciprocity                      & (-0.2639, -0.0532) & (-0.2268, -0.0248) & (-0.1903, 0.008)  & (-0.1487, 0.0359) \\
		totaldegreeSender                & (-1.7193, 11.0226) & (-1.0695, 1.6201)  & (-0.7217, 1.1752) & (-0.4284, 0.7854) \\
		totaldegreeReceiver              & (-3.387, 10.8856)  & (-1.2338, 1.5439)  & (-0.9548, 1.2621) & (-0.6525, 0.9483) \\
		indegreeSender                   & (-4.551, 1.8027)   & (-0.3251, 1.2836)  & (-0.2205, 1.1097) & (-0.0923, 0.8393) \\
		indegreeReceiver                 & (-6.1018, 0.7916)  & (-1.4129, 0.1896)  & (-1.1552, 0.1747) & (-0.8213, 0.1979) \\
		outdegreeSender                  & (-6.2074, 0.4214)  & (-1.2295, 0.3718)  & (-0.905, 0.283)   & (-0.6009, 0.2843) \\
		outdegreeReceiver                & (-4.7354, 2.9554)  & (-0.1644, 1.6278)  & (-0.1012, 1.3676) & (-0.0604, 1.094)  \\
		psABBA                           & (2.0586, 2.4161)   & (2.028, 2.3898)    & (2.0491, 2.3957)  & (2.0709, 2.3846)  \\
		psABBY                           & (-0.1529, 0.2578)  & (-0.1614, 0.2665)  & (-0.1319, 0.2683) & (-0.0798, 0.2405) \\
		psABXA                           & (-0.2021, 0.2382)  & (-0.2035, 0.2223)  & (-0.1854, 0.2056) & (-0.1269, 0.1815) \\
		psABAY                           & (-0.3606, 0.178)   & (-0.3595, 0.1787)  & (-0.3332, 0.1615) & (-0.2548, 0.1026) \\
		\hline
		Number of significant covariates & 3                  & 3                  & 2                 & 2                
	\end{tabular}

\label{apollo_ci}
\caption{Credible intervals for the Apollo 13 data}
\end{table}

\begin{table}[H]
\scriptsize
\centering
\begin{tabular}{l|cccc|cccc}
	& \multicolumn{4}{c|}{First analysis}   & \multicolumn{4}{c}{Extended analysis} \\
	& Flat   & Ridge  & Lasso  & Horseshoe & Flat    & Ridge  & Lasso  & Horseshoe \\
	\hline
	\hline
	inertia                    &          0.255           &           0.222           &           0.196           &             0.104             & 0.231  & 0.15   & 0.113  & 0.094  \\
	reciprocity                &          -0.161          &          -0.122           &          -0.082           &               0               & -0.148 & -0.085 & -0.037 & 0      \\
	totaldegreeSender          &          3.017           &           0.256           &           0.004           &            -0.002             & 1.11   & 0.173  & 0.002  & -0.001 \\
	totaldegreeReceiver        &          1.092           &           0.326           &          -0.004           &             0.004             & 2.007  & 0.291  & 0.002  & 0      \\
	indegreeSender             &          -0.537          &           0.389           &           0.345           &             0.002             & 0.388  & 0.537  & 0.344  & 0.001  \\
	indegreeReceiver           &          -0.905          &          -0.591           &          -0.289           &            -0.002             & -1.007 & -0.092 & -0.003 & 0      \\
	outdegreeSender            &          -1.502          &          -0.319           &          -0.111           &            -0.001             & -0.8   & -0.262 & 0      & -0.001 \\
	outdegreeReceiver          &          0.598           &           0.675           &           0.456           &             0.003             & 0.266  & 0.37   & 0.224  & 0      \\
	psABBA                     &           2.23           &           2.203           &           2.225           &             2.227             & 2.145  & 2.067  & 2.167  & 2.218  \\
	psABBY                     &          0.044           &           0.022           &           0.049           &               0               & -0.073 & -0.086 & -0.001 & 0      \\
	psABXA                     &          0.026           &           0.014           &             0             &               0               & -0.095 & -0.104 & -0.024 & 0      \\
	psABAY                     &          -0.099          &          -0.103           &          -0.052           &             0.001             & -0.232 & -0.302 & -0.123 & -0.001 \\
	cov1                       &                          &                           &                           &                               & -0.035 & -0.023 & -0.009 & 0      \\
	cov2                       &                          &                           &                           &                               & 0.013  & 0.015  & 0.011  & 0      \\
	cov3                       &                          &                           &                           &                               & -0.04  & -0.027 & -0.007 & 0      \\
	cov4                       &                          &                           &                           &                               & 0.023  & 0.014  & 0.012  & 0      \\
	cov5                       &                          &                           &                           &                               & -0.022 & -0.004 & -0.018 & 0      \\
	cov6                       &                          &                           &                           &                               & -0.008 & -0.002 & 0.009  & 0      \\
	cov7                       &                          &                           &                           &                               & -0.021 & -0.014 & -0.011 & 0      \\
	cov8                       &                          &                           &                           &                               & -0.046 & -0.04  & -0.036 & 0      \\
	cov9                       &                          &                           &                           &                               & -0.034 & -0.029 & -0.015 & 0      \\
	cov10                      &                          &                           &                           &                               & -0.02  & -0.024 & -0.02  & 0      \\
	cov11                      &                          &                           &                           &                               & -0.124 & -0.114 & -0.094 & 0      \\
	cov12                      &                          &                           &                           &                               & -0.151 & -0.128 & -0.102 & 0      \\
	cov13                      &                          &                           &                           &                               & -0.003 & -0.024 & -0.01  & 0      \\
	cov14                      &                          &                           &                           &                               & -0.212 & -0.222 & -0.165 & 0      \\
	cov15                      &                          &                           &                           &                               & -0.223 & -0.188 & -0.15  & 0      \\
	cov16                      &                          &                           &                           &                               & -0.126 & -0.131 & -0.123 & 0      \\
	cov17                      &                          &                           &                           &                               & -0.111 & -0.049 & -0.01  & 0      \\
	cov18                      &                          &                           &                           &                               & -0.103 & -0.097 & -0.073 & 0      \\
	cov19                      &                          &                           &                           &                               & -0.096 & -0.124 & -0.039 & 0      \\
	cov20                      &                          &                           &                           &                               & -0.201 & -0.153 & -0.123 & -0.001 \\
	\hline
	\hline
\end{tabular}
\label{mode_apollo}
\caption{Posterior mode values for the Apollo 13 mission analysis}
\end{table}

\newpage
\bibliographystyle{apalike}
\bibliography{ref_for_paper}
\nocite{*}
	
\end{document}